\documentclass{aa}
\usepackage{epsfig}
\usepackage{txfonts}
\usepackage{color}
\usepackage{booktabs}

\def\d{\displaystyle}
\def\kms{km\,s$^{-1}$}
\def\ms{m\,s$^{-1}$}
\def\vs{$v\sin{i}$}
\def\te{$T_{\rm eff}$}
\def\lg{$\log{g}$}
\def\vt{$v_{\rm turb}$}

\def\cd{c\,d$^{-1}$}
\begin{document}

\title{Spectroscopic time-series analysis of R\,Canis Majoris\thanks{Based on observations made with the Mercator Telescope, operated on the island of La Palma by the Flemish Community, at the Spanish Observatorio del Roque de los Muchachos of the Instituto de Astrofisica de Canarias. Table\,1 is only available in electronic form at the CDS via anonymous ftp to cdsarc.u-strasbg.fr (130.79.128.5)
or via http://cdsweb.u-strasbg.fr/cgi-bin/qcat?J/A+A/}.}

\author{H. Lehmann\inst{1}
\and V. Tsymbal\inst{2}
\and F. Pertermann\inst{1}
\and A. Tkachenko\inst{3}
\and D. E. Mkrtichian\inst{4}
\and N. A-thano\inst{4}
}

\institute{Th\"{u}ringer Landessternwarte Tautenburg, Sternwarte 5, 07778 Tautenburg, Germany,
\email{lehm@tls-tautenburg.de, pertermann@tls-tautenburg.de}
\and Crimean Federal University, 295007, Vernadsky av. 4, Simferopol, Crimea, 
\email{vadim.tsymbal@gmail.com}
\and KU Leuven,Instituut voor Sterrenkunde, Celestijnenlaan 200, 3001 Leuven, Belgium,
\email{andrew.tkachenko@kuleuven.be}
\and National Astronomical Research Institute of Thailand, 260 Moo 4, T. Donkaev, A. Maerim, Chiang Mai 50180, Thailand,
\email{davidmkrt@gmail.com; Napaporn.ath@gmail.com};
}


\abstract{R\,Canis Majoris is the prototype of a small group of Algol-type stars showing short orbital periods and low mass ratios. A previous detection of short-term oscillations in its light curve has not yet been confirmed. We investigate a new time series of high-resolution spectra with the aim to derive improved stellar and system parameters, to search for the possible impact of a third component in the observed spectra, to look for indications of activity in the Algol system, and to search for short-term variations in radial velocities.

We  disentangled the composite spectra into the spectra of the binary components.  Then we analysed the resulting high signal-to-noise spectra of both stars. Using a newly developed program code based on an improved method of least-squares deconvolution, we were able to  determine the radial velocities of  both components also during primary eclipse. This allowed us to develop  a better model of the system including the Rossiter--McLaughlin effect and to derive improved orbital parameters. Combining the  results with those from spectrum analysis, we obtain accurate stellar and system parameters. We further deduce at least one oscillation frequency of 21.38\,\cd. It could be detected during primary eclipses only and confirms a previous photometric finding. Results point to an amplitude amplification of non-radial pulsation modes due to the eclipse mapping effect. The presence of a He\,I line in the spectra indicates mass transfer in the R\,CMa system. Calculations of its Roche geometry give evidence that the cool secondary component may fill its Roche lobe. No  evidence of a third body in the system could be found in the observed spectra.}

\keywords {Stars: binaries: eclipsing - Stars: binaries: spectroscopic - Stars: variables: Algols - Stars: oscillations - Methods: data analysis}

\maketitle

\section{Introduction}

R Canis Majoris (R\,CMa, HD\,57167) is a bright ($V$\,=\,5.71), short-period ($P$\,=\,1.13595\,d), Algol-type system. While its secondary component is an Algol-typical cool subgiant (G8\,IV), the  primary is relatively late (F0\,V). The investigation of the R\,CMa system has an extended history. An early review was given by \citet{1946CoPri...22..1W}, a recent comprehensive overview can be found in \citet[hereafter]{2011MNRAS.418.1764B}.

For a long time R\,CMa was the system of lowest total mass and was among the lowest mass ratio/short-period combinations of any known Algol (BB2011). The low mass ratio, overluminosity of the primary, and oversized secondary, also observed in a few other Algol-type systems, caused \citet{1956AnAp...19..298K} to establish a subgroup called R\,CMa-type stars. Later on, the mass ratio and irregularities in the light curve have been controversially discussed, in particular with respect to inaccuracies of many assigned parameter values (e.g. \citealt{1999AJ....117.2980V}, BB2011).

\citet{1985ApJ...297..250T}  was able  to reveal  the secondary component by identifying the sodium doublet in its spectrum, which finally made  a direct determination of the components' masses possible. He obtained masses of 1.1\,M$_\sun$ and 0.17\,M$_\sun$ for the primary and secondary star, respectively. The underlying observational data were relatively poor. The sodium doublet of the secondary is blended with that of the primary in most of the cycle and the number of radial velocity (RV) measurements is very limited,  and with modern observations his results look obsolete. \citet{Glazunova2009} analysed high-resolution spectra from the 2.7m McDonald observatory including the secondary lines and obtained distinctly increased mass values of $M_1$\,=\,1.81\,M$_\sun$ and $M_2$\,=\,0.23\,M$_\sun$. The corresponding mean radii were derived as $R_1$\,=\,1.83\,R$_\sun$ and $R_2$\,=\,1.40\,R$_\sun$, where the value for the primary star is higher than typical for an F0 main sequence star, and both values are higher than the 1.48 and 1.06\,R$_\sun$ cited by \citet{2002AJ....123.2033R}. 

 A recent, comprehensive study of the R\,CMa system is that of BB2011. The authors perform a combined photometric, astrometric, and spectroscopic analysis. The photometry is based on the Hipparcos\,V light curve, U\,B\,V photometry from \citet{1971PASJ...23..335S}, the J and K data of \citet{1999AJ....117.2980V}, and the spectroscopy on high-resolution spectra taken with the HERCULES spectrograph of the Mt. John University Observatory, New Zealand. In the result, the authors derive an improved orbit and absolute parameters of the R\,CMa system. In particular the measured RVs are in good agreement with the results of \citet{Glazunova2009} that had already revised upwards the relatively low masses of \citet{1985ApJ...297..250T}. BB2011 end up with masses of $M_1$\,=\,$1.67\pm 0.08$\,M$_\odot$ and $M_2$\,=\,$0.22\pm 0.07$\,M$_\odot$, resulting in a mass ratio of $0.13\pm 0.05$. The radii are derived as $R_1$\,=\,$1.78\pm 0.03$\,R$_\sun$ and $R_2$\,=\,$1.22\pm 0.07$\,R$_\sun$. From the fit of selected line profiles, the authors find that the projected equatorial rotation velocity of the primary of 82$\pm$3\,\kms\ is in substantial agreement with synchronised rotation. In addition, by adding the Hipparcos positions for the epoch 1991.25 \citet{1997ESASP1200.....E} to those used in \citet{2002AJ....123.2033R}, BB2011 derive narrower constraints on the orbit of a third body in the R\,CMa system (see below).

\citet{2015PKAS...30..231A}, on the other hand, presented the preliminary analysis of medium-resolution spectroscopy of R\,CMa obtained in 2012. They find $M_1$\,=\,$1.83$\,M$_\odot$, $M_2$\,=\,$0.21$\,M$_\odot$, $R_1$\,=\,$1.58$\,R$_\odot$, and $R_2$\,=\,$1.39$\,R$_\odot$, where the mass of the primary and the radius of the secondary are close to the values given by \citet{Glazunova2009}.

Several authors have suggested that R\,CMa is at least a triple system. \citet{1984BASI...12..182R} interpreted the variations seen in the O-C values of the times of minima in terms of the light-travel time (LTT) effect due to the presence of a third body. The authors derive a third-body orbit with $P$\,=\,91.44\,yr, $e$\,=\,0.45, and $\omega$\,=\,25$^\circ$. Later on, \citet{2002AJ....123.2033R} repeated the analysis based on a meanwhile extended set of O-C values, adding the Hipparcos and historic ground-based astrometric data. From the combined analysis they find $P$\,=\,92.8\,yr, $e$\,=\,0.49,  $\omega$\,=\,10.5$^\circ$, and $M_3$\,=\,0.34\,M$_\sun$. The most recent attempt to determine the orbit of the third body is by BB2011. From astrometric analysis they find evidence of a third  body  with a mass of $M_3=0.8$\,M$_\odot$, moving in a $P$\,=\,101.5\,yr, $e$\,=\,0.44, $\omega$\,=\,4$^\circ$ orbit.

The scenario of mass and angular momentum transfer in Algol-type stars  is still up for debate and it is unclear whether the mass transfer can be fully conservative or if systemic mass loss is a general feature of Algols. 
Though there is evidence of non-conservative evolution, the systemic mass-loss rate is poorly constrained by observations. Moreover, systemic mass loss should lead to observational signatures that need to be found. A comprehensive overview on these problems can be found in \citet{2015A&A...577A..55D} where authors compare the mass-loss rates following from different proposed mechanisms of systemic mass loss, favouring the `hotspot' model \citep{2008A&A...487.1129V,2010A&A...510A..13V,2013A&A...557A..40D}
where the mass loss is driven by the radiation pressure of a hotspot. Photoionisation and radiative transfer simulations performed by the authors based on this model lead to effects like IR excesses as observed in some of the known Algols or strong emission lines of highly ionised elements, which are rather untypical for most of the Algol-type systems.

The described problems occur in evolutionary scenarios of Algol-type systems in general and for the  R\,CMa stars, i.e. Algol-type stars with short periods and low mass ratios in particular. Several conservative scenarios have been discussed (see e.g. \citealt{2013A&A...557A..79L}), but  difficulties have been found for the latter group of stars, which would rapidly evolve into contact configurations \citep[][BB2011]{1977ApJ...211..486W,1988AcA....38...89S,1989SSRv...50..205B}. These systems cannot be explained by conservative mass transfer, but result from mass and angular momentum loss from the system in the past, and non-conservative evolution has to be assumed \citep{1985ApJ...297..250T, 2008A&A...486..919M}. BB2011 conclude from their finding of a third component in the R\,CMa system that the problem in evolutionary scenarios connected with the observed low $q,\,P$ combination may be solved, at least for this star, by the devolution of angular momentum to the wider orbit. 

Asteroseismology is an important tool for investigating the interior structure of pulsating stars. It also delivers additional constraints on such fundamental parameters like  masses and radii, in particular for eclipsing binaries with oscillating components, when the results can be combined with those from light curve and spectroscopic analyses. 

$\delta$ Scuti stars are stars with masses of 1.5--2.5\,M$_\odot$ and spectral type A2--F5, showing radial and non-radial p-mode pulsations with frequencies from 4 to 60\,\cd. $\delta$ Scuti-like oscillations have been found in many  EBs (e.g. \citealt{2012MNRAS.422.1250L}) and recently also in several short-period EBs, for example KIC\,9851944 (two pulsating components of equal masses; \citealt{2016ApJ...826...69G}), KIC\,4739791 (rich pulsation spectrum of the primary; \citealt{2016AJ....151...25L}), and KIC\,8262223 (post-mass transfer system; \citealt{2017ApJ...837..114G}).

A special class of stars are known as the oEA stars \citep{2002ASPC..259...96M}. Members are semi-detached, active, eclipsing Algol-type stars showing episodes of mass transfer from a cool (G-K type) donor to a hot (A-F type) gainer star where the latter shows $\delta$ Scuti-type oscillations. These systems can be used to study the interaction between mass transfer episodes, angular momentum exchange, and variable properties of the pulsation modes like amplitude and frequency changes. The best investigated object among the oEA stars is RZ\,Cas \citep[e.g.][]{2004A&A...413..293L,2008A&A...480..247L,2004MNRAS.347.1317R,2006AN....327..905S,2009A&A...504..991T,2018MNRAS.475.4745M}.

\citet{2000IBVS.4836....1M} reported on low-amplitude acoustic mode oscillations in the light curve of R\,CMa with a  frequency of 21.21\,\cd.  Later on, this finding could not be confirmed via short time series of photometric observations \citep{2006Ap&SS.304..169M}. The detection of $\delta$ Scuti-like pulsations in R\,CMa would allow us to apply  asteroseismic methods to its analysis \citep[for applications to $\delta$\,Sct stars  in binary systems see e.g.][]{2014A&A...563A..59M, 2016ApJ...826...69G}. 
This and the finding of a possible impact of mass-transfer effects on the observations would make it a member of the class of oEA stars and an outstanding object for further investigations. We therefore decided to gather time series of medium-resolution spectra of R\,CMa with the eShel and MRES spectrographs attached to the 2.4m Thai National Observatory telescope in 2012 and 2014 and high-resolution spectra with the HERMES spectrograph\footnote{The HERMES spectrograph is supported by the Research Foundation - Flanders (FWO), Belgium; the Research Council of KU Leuven, Belgium; the Fonds National de la Recherche Scientifique (F.R.S.-FNRS), Belgium; the Royal Observatory of Belgium; the Observatoire de Geneve, Switzerland; and the Th\"uringer Landessternwarte Tautenburg, Germany.} at La Palma in January 2016 to search for short-term oscillations in the RVs. The analysis of the medium-resolution spectra gave a hint of an oscillation frequency of 17.38\,\cd\ showing an amplitude of 280\,\ms. The HERMES observations covered one eclipse of the system but no out-of-eclipse phases. In the RV residuals after subtracting the orbit and the distortions due to the Rossiter--McLaughlin effect \citep[][RME hereafter]{1924ApJ....60...15R,1924ApJ....60...22M}
 we also found hints of short-term variations in the RVs of the primary component. To confirm this finding by also investigating the behaviour in out-of-eclipse phases, we repeated the observations in January 2017.
 
In this paper we present the results of the analysis of the high-resolution HERMES spectra. We describe the observations (Sect.\,\ref{Obs}), the decomposition of the observed spectra into the mean spectra of the two components (Sect.\,\ref{decomp}), and the spectroscopic analysis of the decomposed spectra to derive the atmospheric parameters of the components (Sect.\,\ref{SectAna}). In Sect.\,\ref{Impact} we search for the  signature of the third body in our spectra. The determination of the RVs of the stellar components comparing the results from different methods, including a new method of least-squares deconvolution, is described in Sect.\,\ref{SectRVs}. It also lists the derived  orbital solutions. In Sect.\,\ref{Short} we report on the results of our search for short-term variations in the RVs. Section\,\ref{active} is dedicated to the Roche geometry of the system. All results are discussed in Sect.\,\ref{Discuss}, followed by final conclusions.

\section{Observations and data reduction}\label{Obs}

We obtained 169 high-resolution spectra of R\,CMa with the HERMES spectrograph \citep{2011A&A...526A..69R} attached to the Mercator telescope at La Palma in January 2016 and 391 spectra in January 2017. The spectra cover the wavelength range 377--900\,nm having a resolving power of 85\,000. The single exposure time was between 90 and 300\,sec, depending on weather conditions and orbital phase (eclipses), providing a signal-to-noise ratio (S/N) between 60 and 160 (typical value of 100). The spectra were reduced using the standard HERMES pipeline. Subsequently, we normalised the spectra to the local continuum and removed remaining cosmic ray events by using our own routines.

Table\,1 gives the journal of observations together with the measured RVs. 
Table\,1 is only available in electronic form (see the footnote on the title page).

\section{Spectrum decomposition }\label{decomp}

We used the Fourier transform-based KOREL program \citep{1995A&AS..114..393H,Hadrava2006Ap&SS.304..337H,2010ASPC..435...71S} for the decomposition of observed composite spectra. The program determines the optimum shifts that have to be applied to the spectral contributions from the two components to build the mean decomposed spectra of the components, together with the best fitting Keplerian orbit.  We used  the VO-KOREL web service\footnote{https://stelweb.asu.cas.cz/vo-korel/} \citep{2010ASPC..435...71S,2012IAUS..282..403S} to derive the final solution. 
To prepare  the necessary set of input spectra, we implemented an older version of the code kindly provided by P. Hadrava in an ESO-MIDAS \citep{1992ASPC...25..120B} environment, which allowed us
to select the orbital phases and to reject the outliers in a fast and interactive way.

As any Fourier-based method of spectral disentangling, KOREL delivers decomposed spectra that show undulations in their continua. These are understood to be due to indeterminacies in the set of equations (low-frequency Fourier components) in the absence of sufficiently strong time-dependent dilution of spectral lines and/or to imperfect normalisation of the input spectra \citep{2010ASPC..435..207P}. Phase-dependent line strengths of the components reduce these indeterminacies and the undulations can be suppressed remarkably \citep{2008A&A...482.1031H}. In our case, when allowing for timely variable line strengths, KOREL delivered almost constant line strengths for the out-of-eclipse spectra and smooth decomposed spectra with only slight undulations in their continua, which we removed using spline interpolation, taking the complementarity of the undulations between the two components into account. 

We assumed circular orbits and used the values from BB2011 (orbital period $P$ and RV-semiamplitudes $K_1$ and $K_2$, see Table\,\ref{All_RVs}) as starting values. We used the  4960 to 5490\,\AA\ region that is sufficiently extended and contains stronger metal lines from the two components, but no Balmer lines or stronger telluric contributions to determine the orbital parameters of the system. We started with all spectra taken outside primary and secondary eclipses. Then we iteratively excluded all spectra where the deviation of the calculated RVs from the calculated Keplerian orbit exceeds $3\,\sigma$.
 Such deviations mainly occurred for the RVs of the secondary. From an inspection of the corresponding spectra we found that they were the spectra of lowest S/N (below about 80). We ended up with 204 spectra that were used to build the decomposed spectra.  The derived elements of the orbit of the binary were fixed to decompose the regions where broad Balmer lines or telluric contributions occur.

\begin{figure}[h]\centering
\epsfig{figure=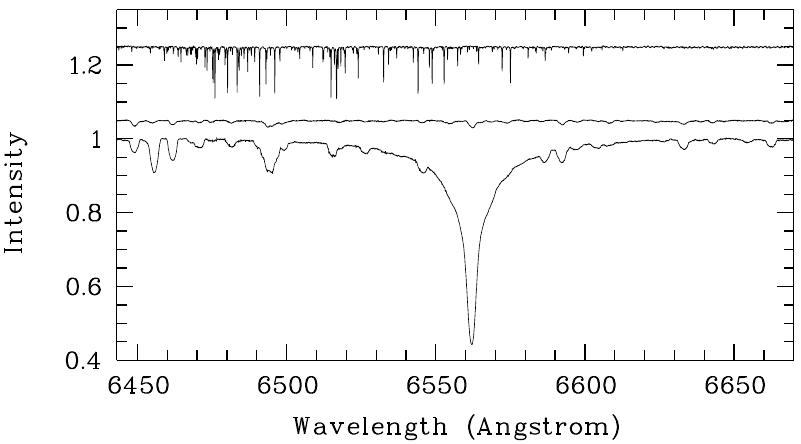, width=9cm, clip=}
\caption{KOREL decomposed spectra of R\,CMa in the H$_\alpha$ region. From bottom to top: Spectra of the primary component,    and (shifted for better visualisation) the secondary component  and the telluric spectrum.}
\label{Halpha}
\end{figure}

The decomposition was done over the spectral range 4000--6800\,\AA, delimited by the lower S/N in the blue range and by the occurrence of strong telluric lines in the red range, using overlapping chunks of 300\,\AA\ width. In those regions where weaker telluric lines occur (e.g.  the H$_\alpha$ region), we included the telluric lines as the spectrum of a third object moving in the virtual orbit of the sun into the KOREL set-up. This method was very successful in separating the telluric lines (Fig.\,\ref{Halpha}), except for the strong telluric bands at 6300\,\AA\ where the weak signal of the secondary did not allow  a decomposition into three components. The deviations between the decomposed spectra in the overlapping regions of neighbouring chunks, where we averaged the results, were marginal.  The resulting decomposed and renormalised spectrum of the primary component had an S/N at 5500\,\AA\ of about 1400, whereas that of the secondary component  had about 50--60, corresponding to the continuum flux ratio between the components at this wavelength of only about 4\% (see bottom panel of Fig.\,\ref{SpecFit2}).

\section{Spectrum analysis}\label{SectAna}
    
The GSSP program \citep{2015A&A...581A.129T} provides a grid search in stellar atmospheric parameters using the spectrum synthesis method.  We used an extended version of the code. First, the continua of the observed spectra are not known a priori  because of strong line blends, in particular in the blue spectral region of cool stars (see  Fig.\,\ref{SpecFit2}). That is why we readjusted them to the continua of the synthetic spectra during the analysis, using spline fits instead of the originally used scaling factors. Second,  we extended the code to  handle three instead of two stellar components to also search for a third body in the R\,CMa system (see Sect.\,\ref{Impact}). Synthetic spectra were computed with SynthV \citep{1996ASPC..108..198T} based on atmosphere models computed with LLmodels \citep{2004A&A...428..993S} for the hot primary star and on MARCS models \citep{2008A&A...486..951G} for the cool companion. Atomic data was taken from the VALD \citep{2000BaltA...9..590K} database. The low temperature of the secondary component required including the molecules into the computation of synthetic spectra, resulting in a corresponding increase in computing time. 

We applied our program to the decomposed spectra in the spectral range 4000\,\AA\ to 6800\,\AA, excluding a short range at 6300\,\AA\ (see Sect.\,\ref{decomp}) and the Mg\,II 4481\,\AA\ line that could not be modelled with the VALD atomic data  at all. For the primary component, the analysis was done iteratively by starting with solar composition, adjusting the atmospheric parameters \te, \lg, \vt, \vs, and [Fe/H], then adjusting the individual abundances of all elements for which we found significant contributions in the spectrum and repeating. The S/N did not allow us to derive individual abundances others than [Fe/H] for the secondary component. During the analysis, the results were coupled via the wavelength dependent continuum flux ratio that was determined from the continuum spectra delivered by SynthV and the radii ratio.  To reduce the degrees of freedom in spectrum analysis, we used fixed values of radii ratio and \lg. For the \lg, we used the values of 4.16 and 3.6 from  BB2011 (errors of 0.03 and 0.1 for the primary and secondary components, respectively, see Table\,\ref{SpecAna}). These values can be considered as the more accurate ones compared to a spectroscopic determination with its interdependence between \lg, \te, and [Fe/H]. The radii and their ratio can than be obtained from \lg\ and the dynamical masses derived from the RVs (Table\,\ref{All_RVs}). From the TODCOR-RVs we obtain a radius ratio of 0.68$\pm$0.09, from LSDbinary 0.67$\pm$0.09, and from BB2011 it follows 0.69$\pm$0.07. We finally fixed the radius ratio to the average value of 0.68.

\begin{table}\centering
\renewcommand{\arraystretch}{1.2}
\caption{Atmospheric parameters of the primary and secondary components derived in this paper and reported in BB2011. Errors are standard deviations, given in parentheses in units of the last digits (here and in the following tables).}
\begin{tabular}{lcccc}
\toprule   
                & \multicolumn{2}{c}{this paper} & \multicolumn{2}{c}{BB2011} \\
                & primary    & secondary         & primary   & secondary      \\
\midrule
\te             & 7033(42)   & 4350(100)         & 7300      & 4350           \\ 
\lg             & 4.2 fixed  & 3.6 fixed         & 4.16(3)   & 3.6(1)         \\
$v_{\rm turb}$  & 2.83(14)   & 2.8(1.0)          &           &                \\
\vs             & 82.2(1.5)  & 64(13)            & 82(3)     & 35             \\ 
Fe/H            & $-$0.07(2) & $-$0.07(20)       &           &                \\
$R_2/R_1$       & \multicolumn{2}{c}{0.68 fixed} & \multicolumn{2}{c}{0.69(3)}\\
\bottomrule
\end{tabular}\label{SpecAna}
\end{table}

\begin{table}\centering
\tabcolsep 1.53mm
\caption{Elemental surface abundances of the primary component of R\,CMa relative to solar values. $N$ is the atomic number of the element. We give the solar abundances for comparison: $A=\log{N_A/N_{\rm total}}$.}\label{Elements}
\begin{tabular}{crrrcrrr}
\toprule
el. & $N$ & \multicolumn{1}{c}{$A_*-A_\odot$} & \multicolumn{1}{c}{$A_\odot$} &
el. & $N$  & \multicolumn{1}{c}{$A_*-A_\odot$}& \multicolumn{1}{c}{$A_\odot$}\\  
\midrule
C   &6 &  $+0.13(18)$ &$-3.65$ &  V   &23  & $+0.22(15)$& $-8.04$\\
O   &8 &  $-0.13(16)$ &$-3.38$ &  Cr  &24  & $ 0.00(06)$& $-6.40$\\
Na  &11&  $+0.33(18)$ &$-5.87$ &  Mn  &25  & $-0.15(12)$& $-6.65$\\
Mg  &12&  $-0.01(09)$ &$-4.51$ &  Fe  &26  & $-0.07(02)$& $-4.59$\\
Si  &14&  $-0.12(14)$ &$-4.53$ &  Co  &27  & $+0.15(26)$& $-7.12$\\
S   &16&  $+0.11(30)$ &$-4.90$ &  Ni  &28  & $-0.09(08)$& $-5.81$\\
Ca  &20&  $-0.05(10)$ &$-5.73$ &  Y   &39  & $-0.26(25)$& $-9.83$\\
Sc  &21&  $+0.19(12)$ &$-8.99$ &  Zr  &40  & $-0.17(30)$& $-9.45$\\
Ti  &22&  $+0.03(06)$ &$-7.14$ &  Ce  &58  & $-0.06(41)$& $-10.46$\\      
\bottomrule                                   
\end{tabular}
\end{table}

\begin{figure}\centering
\epsfig{figure=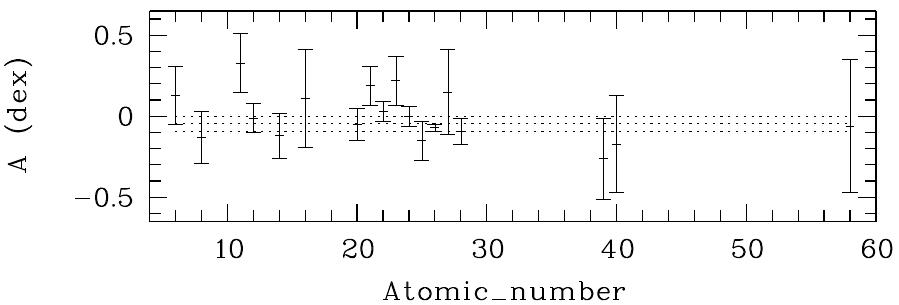, width=9cm, clip=}
\caption{Relative abundances of the primary component vs. atomic number. The dotted lines show the weighted mean and its $1\,\sigma$ error bars.}\label{abundances}
\end{figure}

Table\,\ref{SpecAna} lists the resulting atmospheric parameters of both components. The errors were derived from the full parameter grid so that they already include all interdependencies between the different parameters. Table\,\ref{Elements} lists the elemental surface abundances obtained for the primary component. For the solar values we give the abundances as used by the SynthV program, namely $A=\log(N_A/N_{total})$. They are based on the solar composition given in \citet{2005ASPC..336...25A} and can be converted to the latter values by adding a constant of 12.04.  Figure\,\ref{abundances} shows the relative surface abundances of the primary component versus atomic number of the elements. Abundances of most elements agree within their $1\,\sigma$ error bars with the error-weighted mean of $-0.05\pm 0.05$ derived from all elements. Only for Na, Sc, and V is the abundance  slightly but significantly enhanced.  For the secondary, we could only determine the iron abundance and obtained [Fe/H]=-0.07$\pm$0.20. Irrespective of its large error, the value is in perfect agreement with that obtained for the primary component.

\begin{figure}\centering
\epsfig{figure=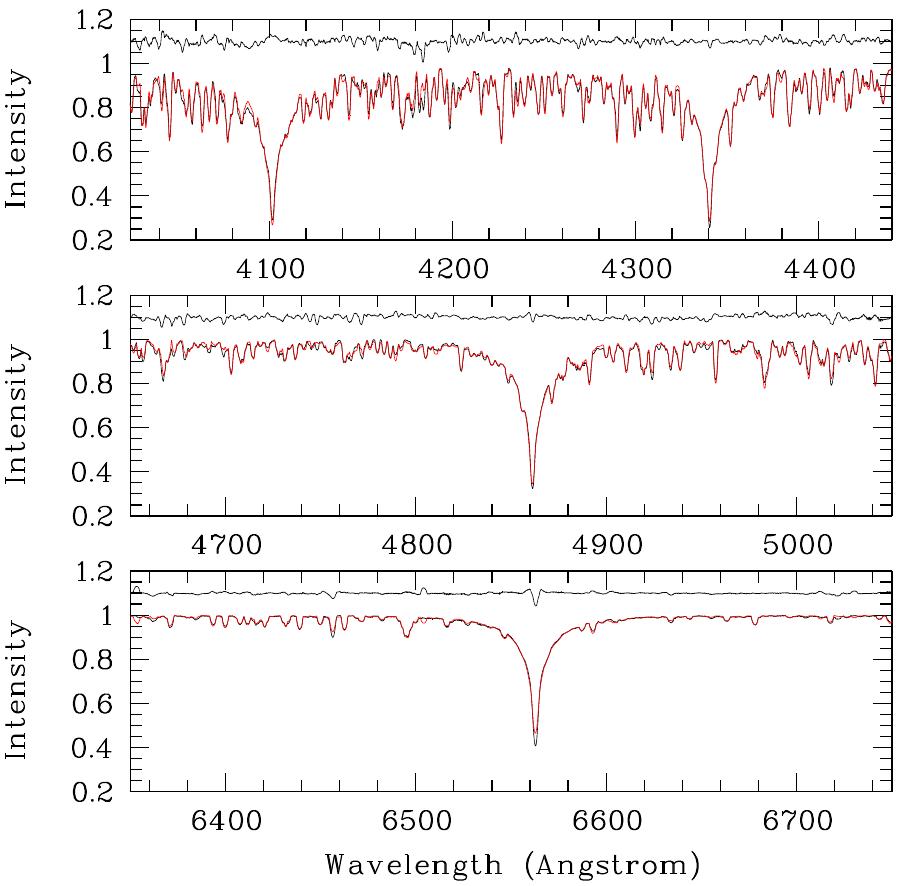, width=9cm, clip=}
\caption{Comparison of the observed renormalised spectrum of the primary component (black) with the best fitting synthetic spectrum (red) in different Balmer lines regions. The difference spectrum is shown in black, shifted by +1.1.}\label{SpecFit1}
\end{figure}

Figures\,\ref{SpecFit1} and \ref{SpecFit2} illustrate the quality of the fit in different wavelength regions. For the primary component (Fig.\,\ref{SpecFit1}), the fit is slightly better  towards longer wavelengths which is due to the wavelength dependent S/N (or instrumental efficiency). For the secondary component the fit is worse in the blue range, which can be explained by the strong wavelength dependence of the flux ratio between the components (Fig.\,\ref{SpecFit2}).

Despite  the high S/N of our decomposed spectra the abundance errors are large  (see Sect.\,\ref{Discuss} for a discussion). We therefore checked our results by building two different composite spectra. In the first case, we coadded the decomposed spectra of the two components, resulting in the mean composite spectrum of high S/N. Since the undulations in the decomposed spectra are complementary they vanish in the coadded spectrum. In  the second case, we used 11 single, observed spectra taken close to the orbital phase of largest separation in RV, rebinned them to the same RV, and coadded them. We modified our version of the GSSP program accordingly and analysed the composite spectra. In both cases we obtained similar values of atmospheric parameters as listed in Table\,\ref{SpecAna}, but with larger error bars (larger by a factor of about 2).  The reason in the first case is that the `decomposition' had to be done again during spectrum analysis, whereas the decomposition with KOREL could make use of the shifts in the spectra of both components introduced by Keplerian motion. In the second case, the larger error arises from the much lower S/N of 11 averaged spectra compared to using the full data set.

\begin{figure}\centering\hspace{1mm}
\epsfig{figure=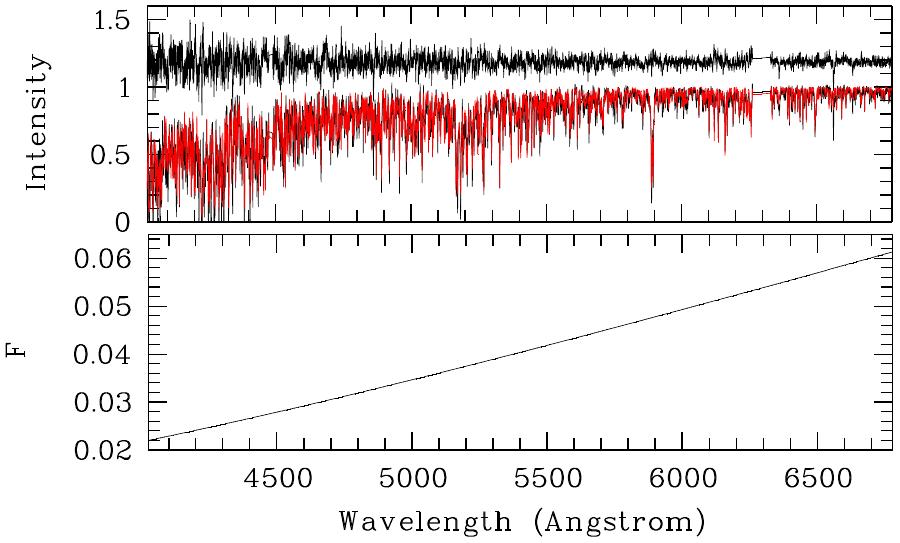, width=8.8cm, clip=}
\caption{Top: As in Fig.\,\ref{SpecFit1}, but for the secondary component and the complete spectral range. Bottom: Continuum flux ratio of the secondary to the primary component.}\label{SpecFit2}
\end{figure}

\section{Missing impact of the third component}\label{Impact}

As mentioned in the introduction, several authors tried to explain the variations observed in the O-C values of times of minima of R\,CMa in terms of the presence of a third body in the system. From the orbital parameters of the third component derived in BB2011 and the spectroscopic mass function, we obtain a maximum RV semi-amplitude $K_3$ of about 6\,\kms\  (for an orbital inclination of 90$^\circ$). Taking the epoch of our observations into account, we expect that RV$_3$ differs from the systemic velocity by not more than about $-1$\,\kms.

\begin{figure}\centering
\epsfig{figure=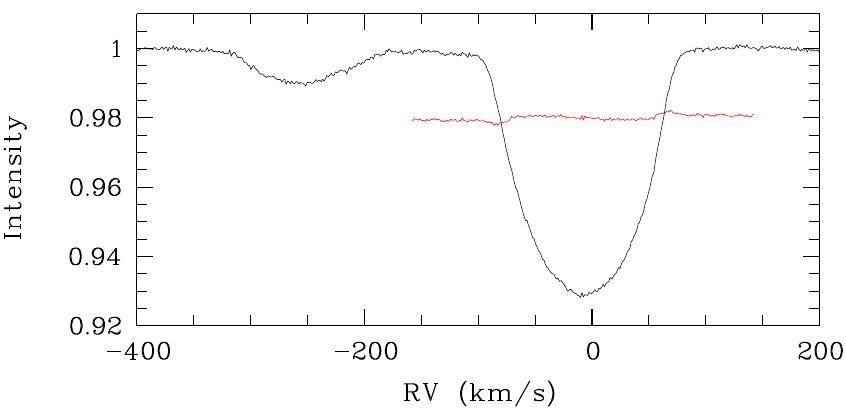, width=8.5cm, clip=}
\caption{Least-squares deconvolution profiles of R\,CMa computed from spectra of the largest RV separation. The red line shows the residuals after subtracting the mirrored profile (shifted by +0.98).}
\label{LSDmaxres}
\end{figure}

Figure\,\ref{LSDmaxres} shows the average profiles of the two stars computed by means of the least-squares deconvolution (LSD) technique (\citealt{1997MNRAS.291..658D}, see also Sect.\,\ref{NewApp}) from 11 coadded spectra taken close to the phase of maximum separation in RV. Profiles were calculated in the 4406--6274\,\AA\ wavelength range, excluding the H$_\beta$ and telluric lines regions. We mirrored the profile of the primary component around its measured RV at $-8$\,\kms\ and subtracted the mirrored profile from the original one. The difference is shown by the red line. We do not see any distortion close to the position of the systemic velocity at $-38$\,\kms, only slight distortions caused by an asymmetry in the outer wings of the profile. From this  we conclude that we cannot find any  evidence of metal lines of the third body in our spectra. 
This also explains why we could not detect a third component with KOREL when trying this by including it in our KOREL set-up  assuming for the third body RVs close to zero (KOREL neglects the systemic velocity).

Finally, we checked  the hypothesis that the third star could be a compact object showing no lines  or only Balmer lines, as  is possible for white dwarfs cooler than about 11\,000\,K. To this end, we included a third component in terms of a veiling caused by the light of a third body in our spectrum analysis. In the result, the error bars of all the parameters of the primary and secondary component, such as  \te, $v_{\rm turb}$, and [Fe/H], increased remarkably compared to the two-component model. In addition, the continuum flux ratio of the third component computed for different combinations of parameters within their $1\,\sigma$ error bars showed positive and negative slopes with wavelength and also positive and negative absolute values. We conclude again that the impact of the third body in our spectra is below the detection limit.

\section{Radial velocities}\label{SectRVs}

Our final goal was to search for low-amplitude short-term variations in the RVs of the primary of R\,CMa possibly related to $\delta$\,Sct-like oscillations. This requires determining highly accurate RVs, in particular in orbital phases during primary eclipse, which  is why we used and compared different methods of RV measurement. The resulting orbits were calculated using the method of differential corrections \citep{1941PNAS...27..175S} for an easy interactive determination of outliers and finally using the PHOEBE program \citep{2005ApJ...628..426P}.

\subsection{Radial velocities from cross-correlation}\label{RVCCF}

In the first step, we determined the RVs of the primary component directly from the observed, composite spectra. For that, we used cross-correlation with the decomposed spectrum of the primary as template. The chosen wavelength range was 4960--5609\,\AA, a region with stronger metal lines but almost no telluric contributions. Spectra were rebinned to  logarithmic wavelength scale and the central parts of calculated cross-correlation functions (CCFs) were fitted by single Gaussian profiles, ignoring the faint contribution of the secondary component.

\begin{figure}\centering
\epsfig{figure=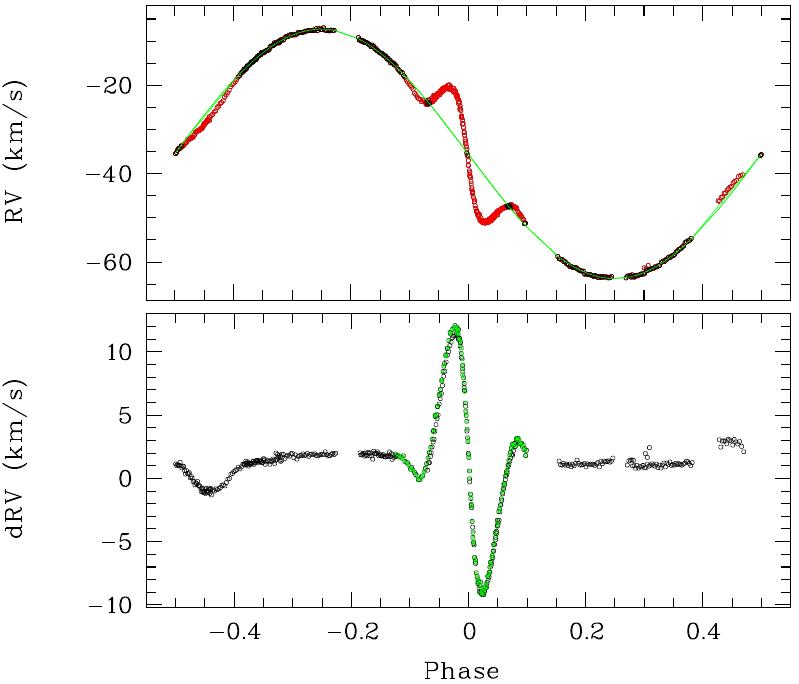, width=8cm, clip=}
\caption{Top: RVs of the primary determined from cross-correlation. It shows the values used for the orbit determination in black, outliers in red, and the resulting orbital curve in green colour. Bottom: RVs of the primary after subtracting the orbital curve.RVs from spectra from 2016 are shown in green colour. }
\label{Orbit_single}
\end{figure}

Figure\,\ref{Orbit_single} shows the obtained RVs together with the O-C values after subtracting the orbit. The data points used for the calculation of the orbit were obtained after iteratively rejecting outliers in the O-C values using 3\,$\sigma$ clipping. In addition to  a pronounced and symmetric RME, we see deviations from the expected behaviour around Min.\,II and at the beginning and end of Min.\,I that may come from the neglected influence of the secondary component. The rms of the scatter in the O-C values in the undisturbed out-of-eclipse parts is of the order of 130\,\ms.

\begin{figure}\centering
\epsfig{figure=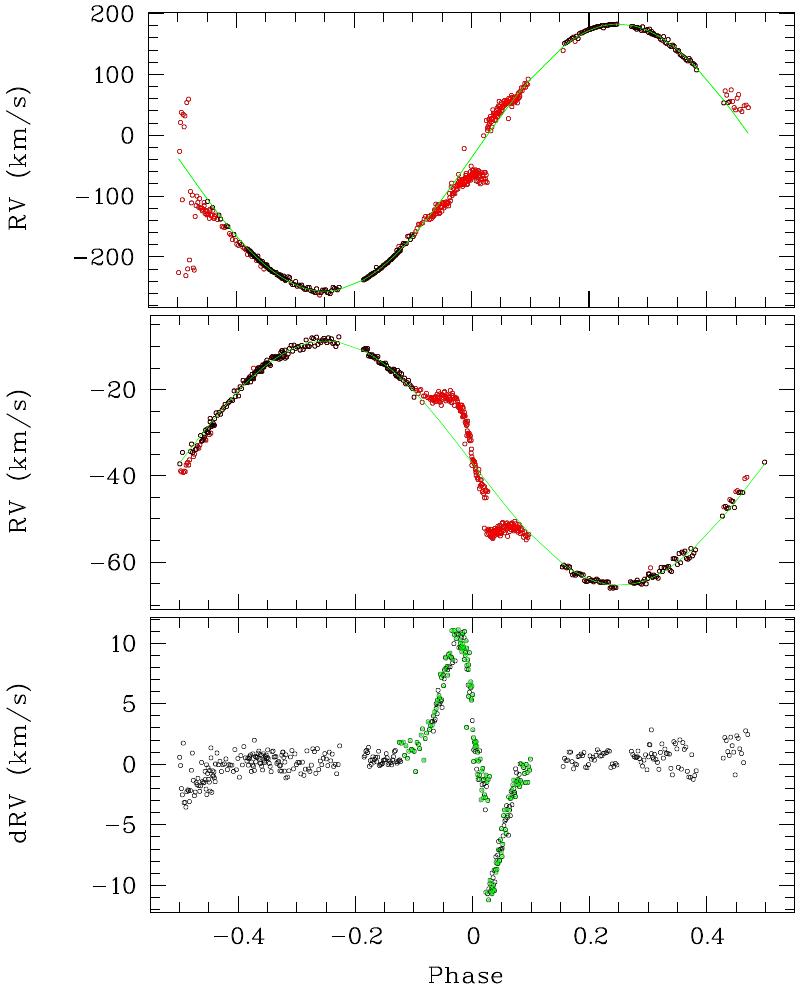, width=8cm, clip=}
\caption{As in  Fig.\,\ref{Orbit_single}, but for the RVs determined with TODCOR for the secondary (top) and the primary (middle panel). Bottom: RVs of the primary after subtracting the orbital curve. The RVs from the spectra from 2016 are shown in green.}
\label{Orbit_TODCOR}
\end{figure}

\subsection{Radial velocities from TODCOR}

We applied an own, Python-based version of the two-dimensional correlation technique TODCOR \citep{1994ApJ...420..806Z}.  TODCOR requires two templates that resemble the observed spectra of the two components as closly as possible. Instead of using synthetic spectra, we used the decomposed spectra of the two components, renormalised according to the flux ratio determined in Sect.\,\ref{SectAna}. TODCOR did not work properly in the blue range of the spectra as used for the cross-correlation method, although metal lines are stronger there. The reason is the faintness of the secondary component and its late spectral type. Its contribution to the composite spectra compared to that of the primary is much stronger in the red part of the spectrum (see bottom panel of Fig.\,\ref{SpecFit2}). We finally used the 6034--6266\,\AA\ range, a region that is almost free of telluric lines and that includes more stronger metallic lines than the regions towards even longer wavelengths. 

Figure\,\ref{Orbit_TODCOR} shows the results. For the calculation of the orbit we excluded all phases around primary eclipse, additionally applying sigma clipping in an iterative way. The rms of the scatter of the orbit-subtracted values (bottom panel of Fig.\,\ref{Orbit_TODCOR}) in out-of-eclipse phases is  about 700\,\ms\  and much higher than for the CCF-based RVs obtained in Sect.\,\ref{RVCCF}. Moreover, the RME of the primary component shows an asymmetry and the RVs of the secondary are strongly disturbed during both eclipses.  A closer look at Fig.\,\ref{Orbit_TODCOR} shows that the asymmetry in the RME of the primary is correlated with the strong deviation of the RVs of the secondary component close to the  centre of primary eclipse.  At these orbital phases, TODCOR could not accurately resolve the 2D-CCFs into the components along the two axes corresponding to the RVs of the two stars, and failed to compute the correct values. On the other hand, the RVs deviations at the primary eclipse ingress and egress, as seen in Fig.\,\ref{Orbit_single},  disappeared. As already mentioned, we assume that this was an effect introduced by neglecting the influence of the secondary in the composite spectra when using cross-correlation to determine the RVs.

\subsection{Radial velocities from LSD profiles:  a new approach}\label{NewApp}

The classical method of least-squares deconvolution  \citep{1997MNRAS.291..658D} is based on using one line mask as template. In the result,  a strong deconvolved line profile is obtained for the star that best matches this template, whereas the contribution of the other component is more or less suppressed. \citet{2013A&A...560A..37T} generalised the method so that it  can simultaneously compute an arbitrary number of LSD profiles from an arbitrary number of line masks. Here we use an approach that makes use of two different line masks and synthetic template spectra for the two stars and delivers two separate LSD profiles, one for each component. In the following, we only give a short description of the method; a more detailed introduction will be published in a forthcoming paper.

The new method first computes the LSD profiles of the two synthetic spectra used as templates and then the recovered profiles from folding the LSD profiles with the used line masks. From a comparison between the recovered and the original spectra, corrections are derived that are used in all the following steps. These corrections stand for the limitations of the classical LSD approach, for example  assuming identical line shapes or linear addition of lines in blends. The LSD profiles computed from the synthetic template spectra as well as the theoretical flux ratio, along with line masks and RV starting values, are used as initial guesses for the final calculations based on time series of observed spectra. Optimised values of the RVs of the components and of the flux ratio of the components are calculated in an iterative way by comparing the composite spectra built from the recovered and the observed spectra. In the result, we obtain time series of separate LSD profiles for the two components. The method offers the possibility of determining accurate RVs from the separated profiles  and simultaneously, which may be the most valuable result for many applications, also of  disentangling the single observed composite spectra to obtain the `recovered' spectra of the components, resulting in a time series of decomposed spectra.

\begin{figure}
\epsfig{figure=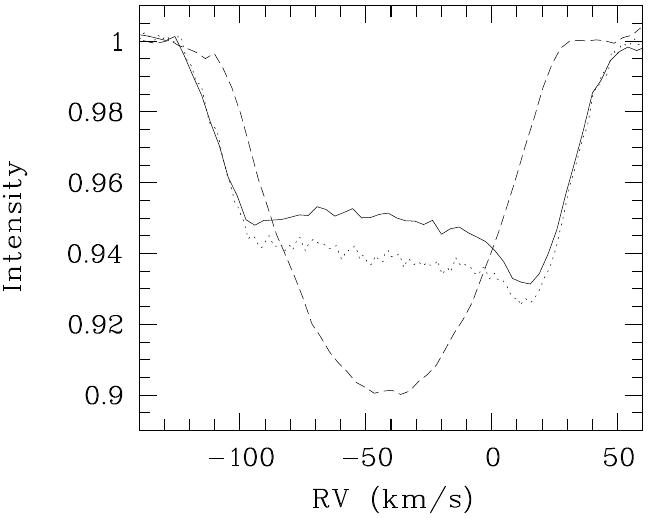, width=6cm, clip=}\centering
\caption{Least-squares
deconvolution profiles from an R\,CMa spectrum taken close to Min\,I computed with LSDbinary for the primary (solid line) and the secondary (dashed line) components, and with the classic LSD approach (dotted line).}
\label{LSD_compare}
\end{figure}

We used the same programs and databases as mentioned in Sect.\,\ref{SectAna} to compute the synthetic spectra and line masks. Figure\,\ref{LSD_compare} shows an example of LSD profiles calculated from one observed composite spectrum taken close to the centre of primary eclipse. Contrary to the classic LSD approach which shows only one profile without giving any information about the secondary component at this orbital phase, we obtain two separate profiles where the profile of the eclipsed star is distorted by the RME.

Least-squares deconvolution profiles were calculated from the same  4960--5609\,\AA\ wavelength
range  used for the calculation of CCFs. Figure\,\ref{Orbit_Phoebe} shows the resulting RVs versus orbital phase. It is based on the solution derived with PHOEBE (see next section). The scatter about the orbital curve of the primary component is about 180\,\ms\ and drastically reduced compared to the results obtained with TODCOR, and only slightly higher than obtained from the CCFs used in Sect.\,\ref{RVCCF}.  In contrast to TODCOR (Fig.\,\ref{Orbit_TODCOR}), the RME shows a symmetric shape because LSDbinary was able to accurately separate the LSD profiles of the two components and to calculate corresponding RVs also in the centre of primary eclipse.

\subsection{Orbital solutions}\label{Orbits}

\begin{figure}\centering
\epsfig{figure=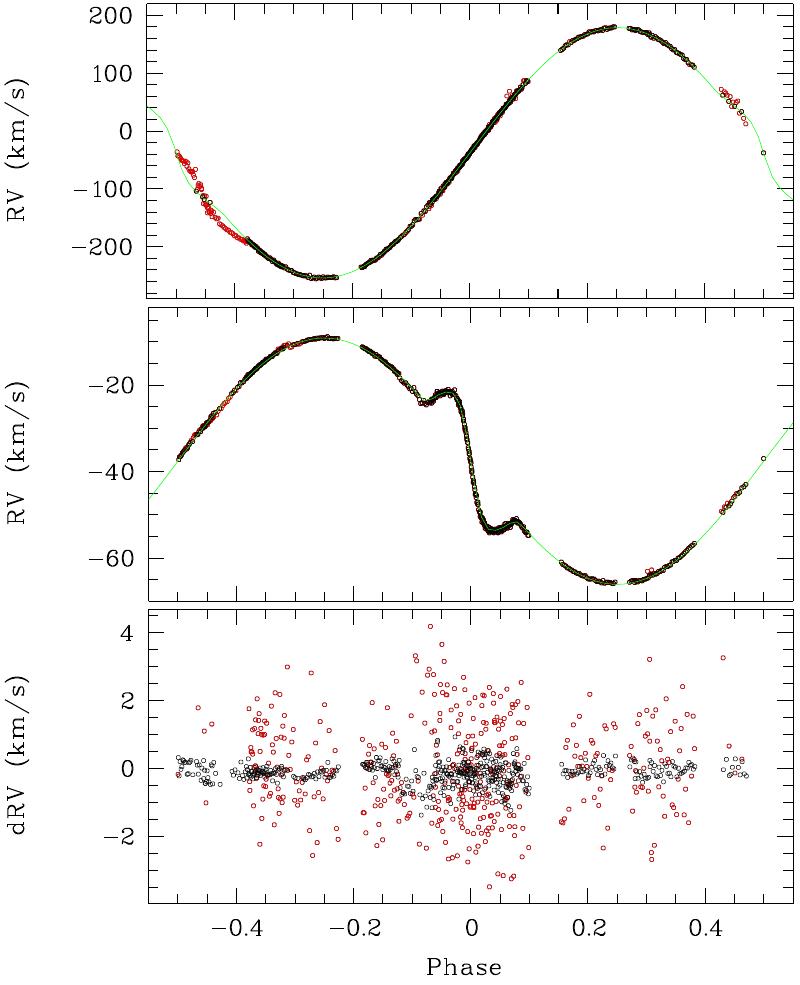, width=8cm, clip=}
\caption{As in Fig.\,\ref{Orbit_single}, but for the RVs fitted with PHOEBE for the secondary (top) and the primary (middle panel). 
Bottom panel: RVs after subtracting the orbital curves for the primary (black) and secondary (red) components.}
\label{Orbit_Phoebe}
\end{figure} 

The quality of the RVs of both components derived with LSDbinary finally allowed us to use PHOEBE to fit the complete orbit including the primary eclipse. PHOEBE models the Roche geometry of the components;    fitting amplitude and shape of the RME during primary eclipse, we also get the inclination $i$ of the orbit and the rotation-to-orbit synchronisation factor $S_1$ of the primary component. For the calculations, we assumed spin-orbit alignment and $S_2$\,=\,1 for the secondary component.

Figure\,\ref{Orbit_Phoebe} shows the fit for the primary component and the O-C residuals. The rms of the scatter about this line is  only 180\,ms$^{-1}$ in out-of-eclipse phases and  400\,ms$^{-1}$ during primary eclipse.  
Table\,\ref{All_RVs} lists the orbital and stellar parameters based on the RVs derived with TODCOR and LSDbinary and compares them with the values given in BB2011. Dependent values were calculated using  the third Keplerian law and the spectroscopic mass function. We also used the orbital inclination determined with PHOEBE to calculate  dynamical masses from the TODCOR solution. Finally, the radii follow from the calculated masses and \lg. As in Sect.\,\ref{SectAna}, the \lg\ values were taken from BB2011.

From the radius of the primary component and its measured \vs\ (Table\,\ref{SpecAna}) we calculate its rotation period to $1.10\pm 0.04$\,d, a value that agrees within the error bars with the assumption of synchronised rotation. PHOEBE, on the other hand, delivers a synchronisation factor of $S_1$\,=\,0.95\,$\pm$\,0.03 corresponding to a rotation period of $1.19\pm0.04$\,d.

\begin{table}\centering
\tabcolsep 2.7mm
\caption{Orbital and stellar parameters derived in this paper and given in BB2011. $T_{\rm MinI}$ is the time of primary minimum $-2\,457\,700$.}\label{All_RVs}
\begin{tabular}{lrrr}
\toprule
                         & TODCOR          & LSDbinary         & BB2011    \\
                         &                 & +PHOEBE           &           \\
\midrule
$P$           (d)            & 1.135938(9) & 1.135956(2)    & 1.13594233\\
$T_{\rm MinI}$           & 0.39217(7)  & 0.39536(6)     & \multicolumn{1}{c}{--}\\
$q$                          & 0.1284(3)   &  0.1294(4)     & 0.129(2)  \\
$v_\gamma$    (\kms)     & $-$37.01(1) &  $-$37.60(2)   & $-$38.7(4)\\
$a$           (R$_\sun$) & 5.629(5)    &  5.655(4)      & 5.66(2)   \\
$K_1$         (\kms)     & 28.24(5)    &  28.5(1)       & 28.6(4)   \\
$K_2$         (\kms)     &  219.9(2)   &  220.4(3)      & 221(3)    \\
$M_1$         (M$_\sun$) & 1.64(7)     &  1.67(1)       & 1.67(8)   \\
$M_2$         (M$_\sun$) & 0.211(9)    &  0.216(2)      & 0.22(7)   \\
$R_1$         (R$_\sun$) & 1.76(7)     &  1.78(6)       & 1.78(3)   \\
$R_2$         (R$_\sun$) & 1.2(1)      &  1.2(1)        & 1.22(3)   \\ 
$i$           ($^\circ$) & \multicolumn{1}{c}{--}& 81.2(1) & 81.7(6)\\
\bottomrule
\end{tabular}
\end{table} 

\section{Search for short-term variations}\label{Short}

We search for short-period oscillations using the O-C values obtained with PHOEBE from the RVs determined with LSDbinary (bottom panel of Fig.\,\ref{Orbit_Phoebe}) that we regard as the most accurate values.  We used the FAMIAS \citep{2008CoAst.157..387Z} and PERIOD04 \citep{2005CoAst.146...53L} programs for the multiple frequency search, together with  successive pre-whitening of the data. In each step when a new frequency was found, all other frequencies were checked frequency by frequency, subtracting the contributions of all other frequencies but the frequency in question, and computing a new periodogram. Finally, all contributions from all frequencies were subtracted and the search for further frequencies in the residuals continued. Following \citet{1993A&A...271..482B}  and \citet{1997A&A...328..544K}, statistically significant peaks in the periodogram had to fulfil the amplitude--S/N\,$\ge$\,4 criterion. The two programs work in a very similar way concerning the frequency analysis of one-dimensional time series, but offer different additional features. Thus, we used FAMIAS to determine the S/N of found frequencies and PERIOD04 to determine the errors via the implemented Monte Carlo simulation.

The rms of the scatter of the O-C values in out-of-eclipse phases based on the LSDbinary--RVs is of only 150\,\ms\ and distinctly lower than the amplitudes of the previously found short-term variations mentioned in before. In consequence, we could not detect any significant frequency. The largest peak in the periodogram found at 2.27\,\cd\ had an amplitude of 64\,\ms\ and an S/N of only 3.5.

The scatter of the O-C values during primary eclipse is much larger than in out-of-eclipse phases (see Fig.\,\ref{Orbit_Phoebe}). This could be caused by an imperfect modelling of the RME by the PHOEBE program and/or to less precise RVs determined from the LSD profiles disturbed by the RME. However, it could also be caused by an amplification of observed oscillations from non-radial pulsation modes  of special $l,m$ combinations due to the eclipse mapping effect.  This effect, also called the spatial filtration effect, was described by  \citet{2003ASPC..292..369G} and \citet{2005ASPC..333..197M} as a tool for the identification of non-radial pulsation modes and applied to the oEA stars AB\,Cas  \citep{2004MNRAS.353..310R} and RZ\,Cas  \citep{2003ASPC..292..369G,2018MNRAS.475.4745M}. It is based on the fact that the surface of the pulsating component is scanned during primary eclipse by its companion which works as a timely variable spatial filter causing specific changes in the pulsation light curve and also in the observed line profiles. Changes in RV and the resulting pulsation amplitude amplification during primary eclipse in dependence on the $l,m$ numbers of the pulsation modes were calculated in detail by \citet{2010Tkachenko}. For an orbital inclination close to 90$^\circ$ and special modes like $(l,m)$\,=\,(1,0), (2,1), (3,2), or (3,3), the observed amplitudes can be enhanced by a factor of more than four.
 The eclipse mapping effect can also be used to reconstruct the oscillation pattern on the surface of pulsating stars in eclipsing binaries from the observed light curves using the dynamic eclipse mapping technique as introduced by \citet{2011MNRAS.416.1601B}.

\begin{figure}\centering\hspace{-1mm}
\epsfig{figure=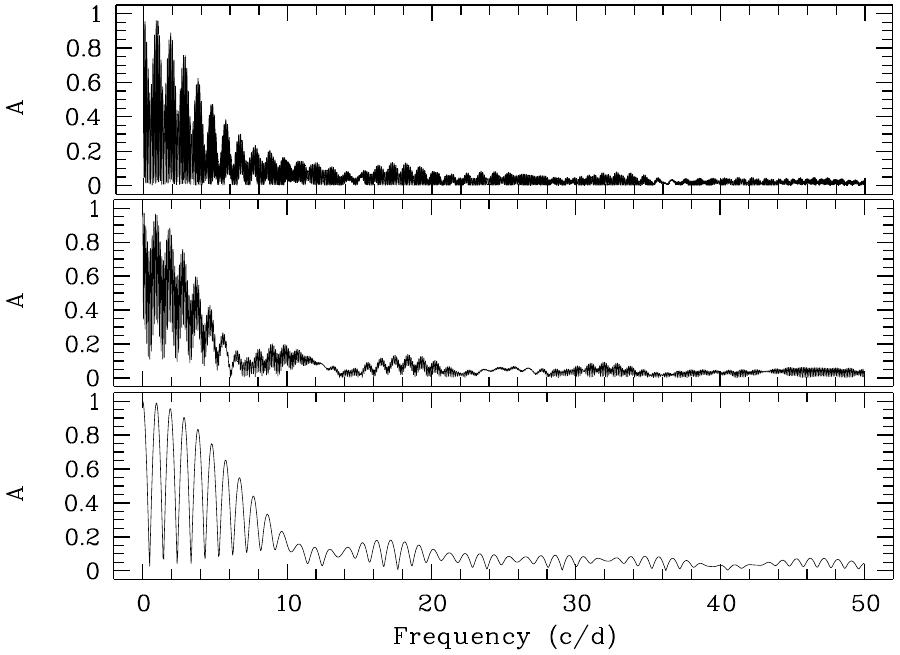, width=9cm, clip=}
\caption{Spectral window functions of in-eclipse data. From top to bottom for  all data, and for data from 2016 and 2017.}
\label{Window}
\end{figure}

For the analysis of the primary eclipse phases we had 155 data points from spectra observed in three nights in 2016 and 82 data points from two nights in 2017 at our disposal. The window functions of the two data sets are shown in Fig.\,\ref{Window}. It can be seen that the frequency resolution obtained from the 2016 data is distinctly better than from 2017. In both cases a strong one-day aliasing occurs. 

The poor window function of the data from 2017 did not allow  a reliable multiple frequency search.  The strongest peak in the periodogram was found at 0.49\,\cd. The clumpy phase diagram showed that this is an alias frequency caused by the time sampling. The least-squares fit of PERIOD04 did not converge when adding further frequencies found in the residuals after subtracting the first one. In the following we present the results from investigating the in-eclipse data from 2016 and from the combined data obtained in 2016 and 2017.

\begin{table}\centering
\caption{Frequencies found in the in-eclipse O-C values.}\label{OsciFreq} 
\begin{tabular}{lrllr}
\toprule
& frequency & amplitude & \multicolumn{1}{c}{phase} & \multicolumn{1}{c}{S/N}\\
& \multicolumn{1}{c}{(\cd)}     & \multicolumn{1}{c}{(\kms)}    & \multicolumn{1}{c}{($2\pi$)}\\
\midrule
\multicolumn{5}{c}{data from 2016}\\
$f_1$ &  8.9(3)    &0.26(3) & 0.6(3)  &16.3\\
$f_2$ & 21.383(5)  &0.22(3) & 0.21(2) & 9.9\\
$f_3$ & 14.0(2)    &0.16(4) & 0.2(2)  & 7.7\\
$f_4$ & 31.8(7)    &0.14(3) & 0.5(3)  &19.0\\
\midrule
\multicolumn{5}{c}{all data}\\
$F_1$ &  2.6845(2) &0.19(3) & 0.61(3) & 5.8\\
$F_2$ & 21.3832(1) &0.23(3) & 0.29(2) &13.1\\
$F_3$ &  8.683(5)  &0.21(3) & 0.1(2)  & 7.2\\
$F_4$ & 30.4783(2) &0.13(3) & 0.97(3) & 4.5\\
$F_5$ & 65.7902(4) &0.12(3) & 0.35(4) & 3.8\\
\bottomrule
\end{tabular}
\end{table}

\begin{figure*}\centering
\epsfig{figure=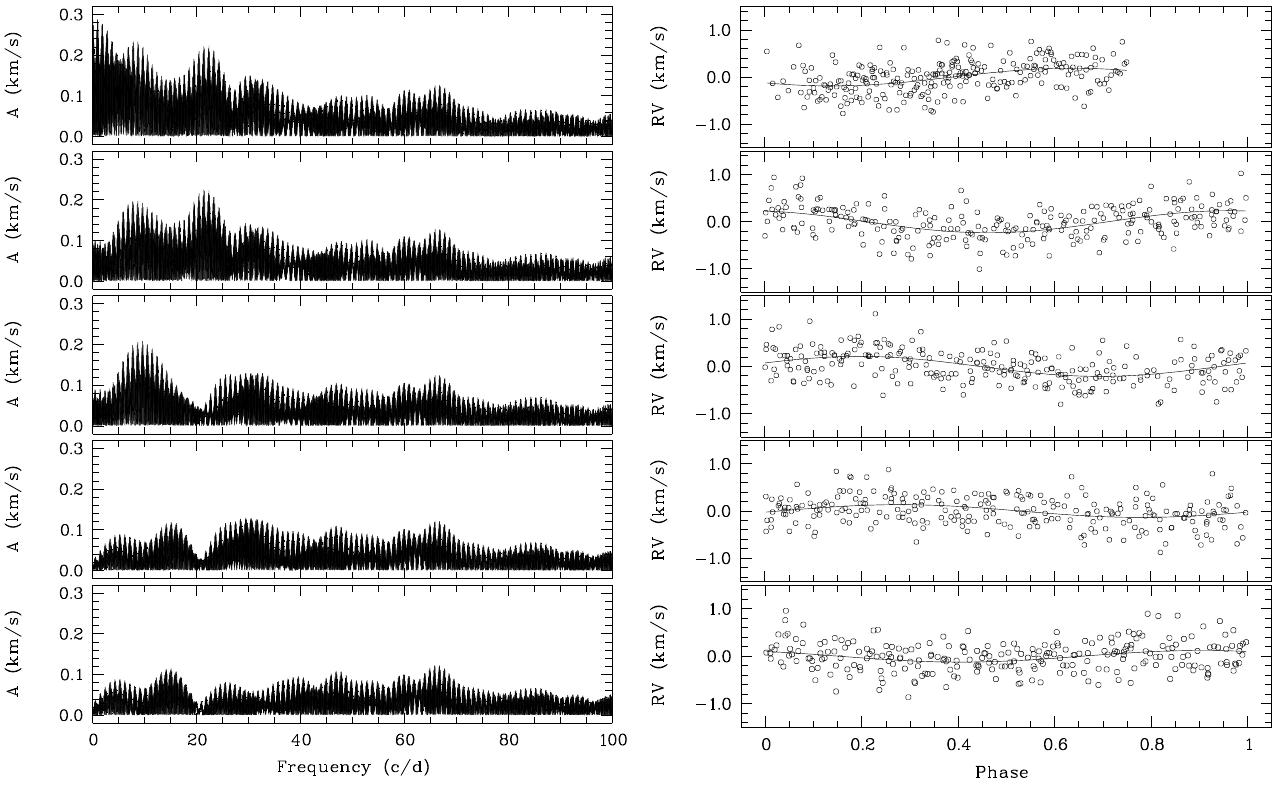, width=16cm, clip=}
\caption{Left: Fourier-amplitude periodograms from all in-eclipse data obtained in different steps of successive prewhitening, showing the highest peaks for frequencies from $F_1$ to $F_5$ (from top to bottom). Right: Corresponding contributions of $F_1$ to $F_5$ after subtracting the contributions of all other frequencies in each case. The solid lines were calculated from the frequencies, amplitudes, and phases listed in Table\,\ref{OsciFreq}.}
\label{FFTall}
\end{figure*}
Table\,\ref{OsciFreq} lists the results from 2016 in the upper part, and from all in-eclipse data in the lower part. The numbering of frequencies corresponds to the order in which they were found. Parameter errors were computed using the Monte Carlo simulation implemented in PERIOD04 based on 10\,000 trials in each case. Figure\,\ref{FFTall} shows the periodograms obtained in different steps of pre-whitening and the corresponding phase plots folded with the found frequency after subtracting the contributions of all other frequencies.
 
The agreement between $f_2$ and $F_2$ is excellent. Frequencies $f_1$ and $F_3$ agree within their error bars, the same for $f_4$ and $F_4$ when taking the one-day aliasing into account. Frequency $F_5$ is not a significant finding according to the S/N>4 criterion, we list it only as the largest peak occurring in the residuals after subtracting $F_1$ to $F_4$. Frequency $F_1$ cannot be found in the spectra from 2016. Data in the corresponding phase plot (Fig.\,\ref{FFTall}) does not fill the full phase range, and we assume that it could be an alias frequency created by the time sampling. We end up with one secure detection, $F_2$, which had already been found in the light curve of the binary system \citep{2000IBVS.4836....1M}. It is present in both the data from 2016 and in all data and shows  small errors in frequency and phase in both
data sets. Furthermore, we found two frequencies, $F_3$ and $F_4$, which need additional confirmation. Both show large errors in frequency and phase in the data from 2016, and $F_3$ also in phase in the complete data set.

PERIOD04 provides the possibility of checking for amplitude and phase variations between two subsets. We applied this to the combined data set, separating it into data from 2016 and 2017. The variations in amplitude and phase for all three frequencies, $F_2$ to $F_4$, were marginal. Thus, we assume that they are present in the single data set from 2016 as well, and that their detection was prevented by the poor window function.

The rms of the residuals after subtracting $F_1$ to $F_5$ is  296\,\ms\ against 400\,\ms\  before subtracting, corresponding to a reduction in the sum of squares of 45\%.  Nevertheless, it is about twice the value obtained from the out-of-eclipse data. Though we achieved distinctly better defined RVs during the eclipses with our new LSDbinary program compared to other methods, it is obvious that our in-eclipse RVs are still less accurate than those obtained from the out-of-eclipse phases. Moreover, the results of frequency search strongly depend on the accuracy of the obtained PHOEBE solution, i.e.  on the quality of the fit applied to the distortion in the orbital RV curve caused by the RME. This is why the errors following from the Monte Carlo simulation (Table\,\ref{OsciFreq}) underestimate the true errors, and we limit the number of digits of the three frequencies found to give 21.38, 8.68, and 30.48\,\cd.

\section{R\,CMa:  an active binary system?}\label{active}

\begin{figure}\centering
\epsfig{figure=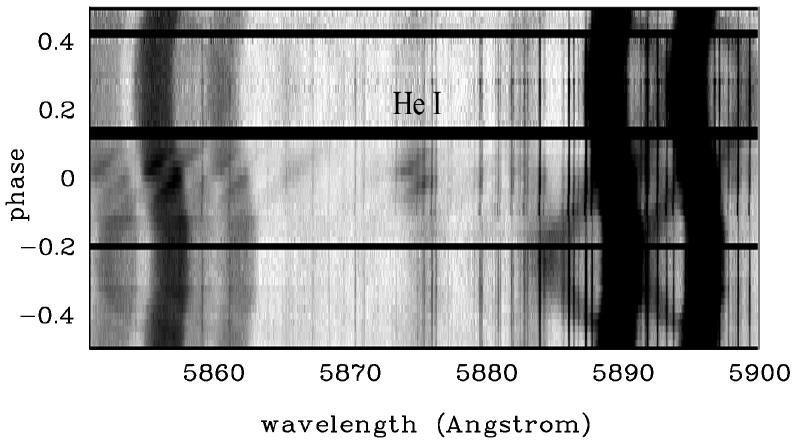, width=8cm, height=4.5cm}
\caption{Orbital-phase binned spectra of R\,CMa indicating the presence of the He\,I line at 5876\,\AA. The straight and sharp vertical lines are of telluric origin.}
\label{HeI}
\end{figure}

We detected the He\,I 5876\,\AA\ line in our medium-resolution  spectra from 2012 and 2014, showing variable line strengths, and in some of our high-resolution HERMES spectra taken in 2016 and 2017. Figure\,\ref{HeI} shows the 560 HERMES spectra averaged into 50 bins evenly distributed over orbital phase. It can be seen that the signal of the He\,I line is very weak, but it is clearly detectable around zero phase where the line depths are enhanced by the RME.

The He line indicates the presence of some hot regions, for example  a hot spot on the primary caused by mass transfer. We therefore investigated the Roche geometry of the system a bit more closely. 
The modified Roche potential of a binary in a circular orbit \citep[see e.g.][]{1979ApJ...234.1054W, 2016ApJS..227...29P} can be written in Cartesian coordinates  as

\begin{equation}
\begin{array}{l}
\Omega_x = \frac{\d 1}{\d x}+q\left(\frac{\d 1}{\d 1-x}-x\right)
          +S^2\,(1+q)\,\frac{\d x^2}{\d 2}\\
\Omega_y = \frac{\d 1}{\d y}+\frac{\d q}{\d 1+y^2}+S^2\,(1+q)\,\frac{\d y^2}{\d 2}\\
\Omega_z = \frac{\d 1}{\d z}+\frac{\d q}{\d\sqrt{1+z^2}} 
\end{array}
\label{Roche_eq}
,\end{equation}
where $x,y,z$ are in units of the separation $a$ between the components and centred at the position of the star in question. Coordinate $x$ points towards the other component, $y$ lies in the orbital plane, and $z$ is perpendicular to the orbital plane. In our case, we have to place the secondary component at the point of origin and to use the inverse of the mass ratio used so far, i.e. $q$\,$=$\,$M_1/M_2$. The parameter  $S$ is the rotation/orbit synchronisation factor.

\begin{figure}\centering
\epsfig{figure=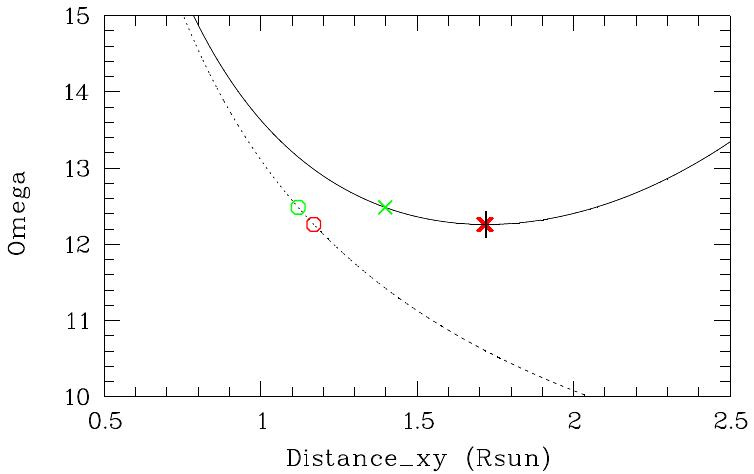, width=7.5cm}
\caption{Modified Roche potential vs. distance from the centre of the secondary star in the $x$-direction (solid line) and $y$-direction (dotted line). The circles and crosses indicate the positions for stars of different radii (see text).}
\label{Roche_xy}
\end{figure} 

Figure\,\ref{Roche_xy} illustrates an example. The Roche potential  $\Omega_x$ is shown  by the solid line, calculated for $a$\,=\,5.673\,R$_\sun$, $q$\,=\,0.1321, and $S$\,=\,1 (assuming synchronised rotation for the secondary star). The minimum of $\Omega_x$ (the sign of the modified potential is opposite the sign of  its physical definition) gives the position of the inner Lagrangian point at $L_1$\,=\,1.72\,R$_\sun$, indicated by the black plus sign; $\Omega_y$ is shown  by the dotted line; the green circle represents the value of $\Omega_y$ for a radius of the secondary component in $y$-direction of $R_y$\,=\,1.12\,R$_\sun$; and the green cross the corresponding value of $\Omega_x$. The x-coordinate of this cross gives the distance of the substellar point of the secondary star from its centre. It does not reach $L_1$, and the star does not fill its Roche lobe. The red circle, on the other hand, gives the radius in the y-direction for the case that the substellar point coincides with $L_1$ and the secondary component fills its Roche lobe. The critical value is $R_y$\,=\,1.170\,R$_\sun$.

In our case, the PHOEBE solution delivers the surface potentials of the primary and secondary components of $\Omega_1$\,=\,21.1$\pm 0.2$ and $\Omega_2$\,=\,12.81$\pm 0.05$, respectively. The calculated potential in $L_1$ is 12.48, slightly below $\Omega_2$,  and the potential in $L_2$ is 11.83. According to this solution, the secondary star does not fill its Roche lobe, but is very close to filling it.  The Roche-lobe filling factor,  defined in \citet{1984ApJS...55..551M} as the ratio of the surface potential to the potential in $L_1$, is 0.974$\pm$0.004.

Our radii given in Table\,\ref{All_RVs} are based on the spectroscopically determined  masses and the \lg\ taken from light curve analysis. Their relatively large errors do not allow us to determine whether  the secondary component fills its Roche lobe.  The radius of the secondary photometrically derived by BB2011, on the other hand, should be very close to the definition of $R_y$ and its value of 1.22$\pm$0.03  corresponds to a surface potential of about 12.23 which lies between the potentials in $L_1$ and $L_2$. This and the presence of the He\,I line in the spectra indicates that the secondary fills its Roche lobe.
  
\section{Discussion and conclusions}\label{Discuss}

 Whereas most of the derived atmospheric parameters agree well with previous findings, our \te\ of the primary star of 7033$\pm$44\,K is distinctly lower. \citet{1996ApJ...458..371S} derived 7310$\pm$100\,K, based on the observed B$-$V value of +0.27 obtained from the light curves of \citet{1985BASI...13..261R}, and assumed the primary component of R\,CMa as an F1 star.
This value of 7310\,K was adopted later by several authors like \citet{1999AJ....117.2980V} or \citet{2006Ap&SS.306..159M} in their light-curve analyses and also by BB2011. 

\begin{figure}\centering
\epsfig{figure=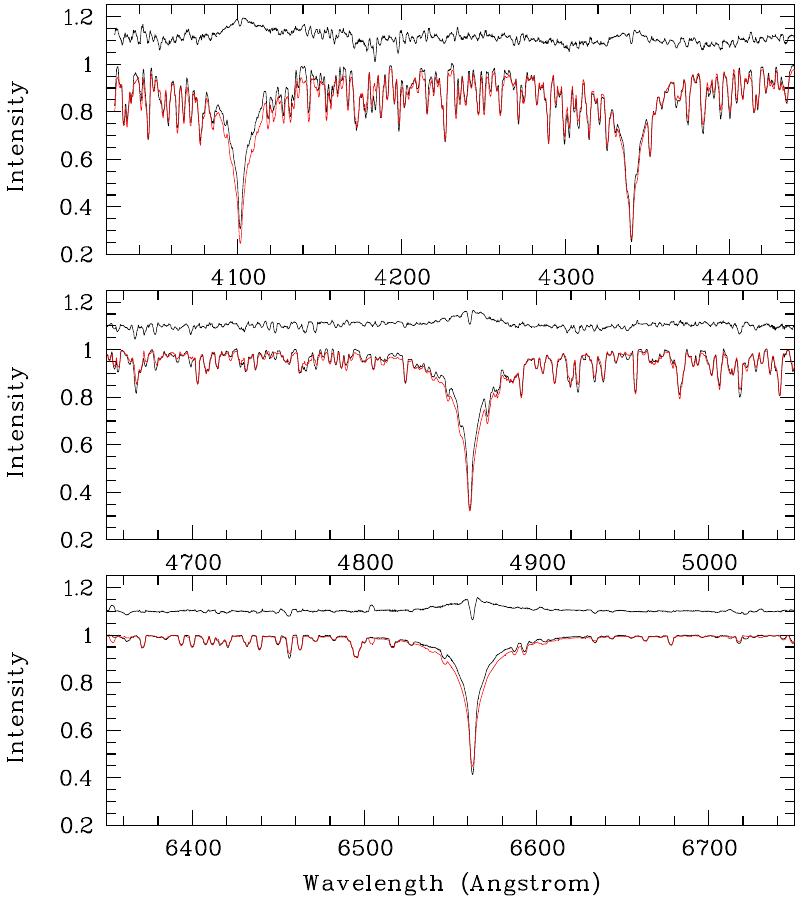, width=8cm}
\caption{As in Fig. 3, but for the best fit using \te=7500\,K.}
\label{test7300}
\end{figure} 

We repeated the spectrum analysis with \te\ fixed to 7300\,K using [Fe/H] as a free parameter. In the result, the relative surface iron abundance increases from $-0.07$ in the case of \te=7033\,K to $-0.01$. Figure\,\ref{test7300} shows the quality of the fit. The rms of the O-C residuals is  39\% higher than for 7300\,K. From this and the comparison of Fig.\,\ref{test7300} with Fig.\,\ref{SpecFit1} we conclude that our \te\ is much more reliable than the previously assumed value.

\begin{table}\centering
\caption{Elemental surface abundances derived in the present study and by TL1989.}\label{Tomkin} 
\begin{tabular}{ccc}
\toprule
element & present study & TL1989\\
\midrule
C  & $~~~0.1 \pm 0.2$ & $0.0\pm 0.2 $\\
O  & $-0.1 \pm 0.2$ & $0.3\pm 0.3 $\\
S  & $~~~0.1 \pm 0.3$ & $0.1\pm 0.2 $\\
Fe & $-0.07\pm 0.02$& $0.1\pm 0.1 $\\
\bottomrule
\end{tabular}
\end{table}

Table\,\ref{Tomkin} compares our elemental surface abundances of the primary of R\,CMa with those obtained by \citet[][hereafter TL1989]{1989MNRAS.241..777T} from Reticon spectra taken with the 2.7m telescope of the McDonald observatory. These spectra had a S/N between 200 and 500. Authors used the equivalent width method on selected spectral lines for the abundance analysis. Despite  the higher S/N of our decomposed spectrum, our abundance errors are relatively large and comparable with those from TL1989. 
We assume that  the S/N is not the limiting factor in our analysis, but the accuracy of spectrum normalisation and of the VALD atomic data, and the interdependence between the surface iron abundance and other atmospheric parameters like \te, \lg, and $v_{\rm turb}$. Moreover, we derived the errors from the full grid in all atmospheric parameters so that they include all interdependencies, whereas TL1989 used fixed values of \te, \lg, and $v_{\rm turb}$. 
Our C, O, and S abundances agree with those from TL1989 within the error bars, but our iron abundance is distinctly lower. The main reason for  the difference is the strong interdependence between [Fe/H] and \te\ as shown in  the previous paragraph. We derived \te\ of 7033\,K, whereas TL1989 adopted 7500\,K, based on the B$-$V and H$_\beta$ indices from \citet{1977AJ.....82...51G}. 

Our value of 64$\pm$13\,\kms\ for the \vs\ of the cool companion is not very precise, which is due to its faintness and the mentioned interdependence between the parameters. It agrees within the $1\,\sigma$ error bars with the value of 54.4$\pm$1.3\,\kms\ expected for the case of synchronised rotation. 

Comparing the stellar and system parameters derived with LSDbinary/PHOEBE with those given in BB2011 (Table\,\ref{All_RVs}) we see that most values agree well;   except for the stellar radii and the not iterated \lg, the error bars of all parameters determined in our study are distinctly smaller. For the slightly variable orbital period we have to take into account  the epoch difference. From the O-C diagram of times of minima of R\,CMa, for example provided by the O-C gateway\footnote{http://var2.astro.cz/ocgate/} \citep{2006OEJV...23...13P}, we see that our value of $T_{\rm MinI}$ perfectly fits into the expected course. 

The finding of oscillation frequencies in the measured RVs was not straightforward. Significant peaks in the periodograms were only found in the data taken in eclipse phases in 2016 and could not be detected in the out-of-eclipse phases  or in the in-eclipse data from 2017.  Reasons including the amplitude amplification of non-radial pulsation modes during the eclipses have been discussed in Sect.\,\ref{Short}.
We consider the frequency of 21.38\,\cd\ as a secure finding. This is supported by the fact that the frequency value is close to the 21.21\,\cd\  found by \citet{2000IBVS.4836....1M} (who
give $\pm 1$\,\cd\ due to aliasing) in the out-of-eclipse light curve of R\,CMa, which had an amplitude of only 4.4\,mmag on average.  The other two frequencies of 8.68 and 30.48\,\cd\ show larger errors and deviations between the two investigated data sets in frequency and phase and need additional confirmation from independent observations.
This study also shows that the disentangling of the RV variations caused by the RME and by pulsation requires a highly accurate modelling of the RME.  The frequencies found in the residuals strongly depend on the achieved accuracy of modelling and thus the frequency errors following from frequency analysis alone may underestimate the true errors.

The presence of the He\,I line points to mass transfer in the R\,CMa system. Based on a Roche model of the system and the radius of the secondary star measured from the light curve, the cool companion fills its Roche lobe.  The surface Roche potential of the secondary star obtained with PHOEBE, on the other hand, gives a Roche lobe filling factor of 0.97 close to unity. The applied Roche model is a relatively simple approach that assumes that the masses of the two stars are so concentrated that their combined gravitational potential can be approximated by the potentials of two point masses.  This is why we can assume, despite the small deviation from unity of the derived filling factor, that the star undergoes at least temporary phases of mass transfer.

The occurrence of mass exchange indicated by the presence of the He\,I line makes R\,CMa a member of the class of oEA stars as defined by \citet{2002ASPC..259...96M}, namely active (showing mass-transfer episodes) eclipsing Algol-type stars where the gainer (the hot primary component) shows $\delta$ Scuti-like oscillations and makes the star a very interesting object for further investigations.

We could not find any  signature of a third component in the R\,CMa system in our spectra  from using KOREL for spectral disentangling,  from spectrum analysis, or in the computed LSD profiles.

Based on the RVs calculated with our new program LSDbinary we could establish an accurate orbital solution and model the RME during primary eclipse. LSDbinary was not developed in particular to determine RVs, however. Its main purpose is to compute separated LSD profiles of the components of binary systems from composite spectra and to use them to calculate the single spectra of the components from folding the LSD profiles with the used line masks. In this way, spectral disentangling of single spectra in a time series can be performed. This and the fact that the decomposed LSD profiles  reflect the distortions caused not only by the RME, but also those from pulsation can open new horizons in asteroseismic investigations.

\begin{acknowledgements} 
This research was accomplished with the help of the VO-KOREL web service, developed at the Astronomical Institute of the Academy of Sciences of the Czech Republic in the framework of the Czech Virtual Observatory (CZVO) by P. Skoda and J. Fuchs using the Fourier disentangling code KOREL deviced by P. Hadrava. Frequency analysis was partially obtained with the software package FAMIAS developed in the framework of the FP6 European Coordination Action HELAS. HL and FP acknowledge support from DFG grant LE 1102/3-1; VT acknowledges support from RFBR grant No. 15-52-12371. The research leading to these results has  received partial funding from the European Research Council (ERC) under the European Union's Horizon 2020 research and innovation programme (grant agreement N$^{\circ}$670519: MAMSIE), from the Belgian Science Policy Office (Belspo) under ESA/PRODEX grant ``PLATO mission development'', and from the Fonds Wetenschappelijk Onderzoek-Vlaanderen (FWO) under  grant agreement G0H5416N (ERC Opvangproject).
\end{acknowledgements} 

\bibliographystyle{aa}
\bibliography{AA2016_29914_final} 

\begin{thebibliography}{73}
\expandafter\ifx\csname natexlab\endcsname\relax\def\natexlab#1{#1}\fi

\bibitem[{{A-Thano} {et~al.}(2015){A-Thano}, {Mkrtichian}, \&
  {Komonjinda}}]{2015PKAS...30..231A}
{A-Thano}, N., {Mkrtichian}, D.~E., \& {Komonjinda}, S. 2015, Publication of
  Korean Astronomical Society, 30, 231

\bibitem[{{Asplund} {et~al.}(2005){Asplund}, {Grevesse}, \&
  {Sauval}}]{2005ASPC..336...25A}
{Asplund}, M., {Grevesse}, N., \& {Sauval}, A.~J. 2005, in Astronomical Society
  of the Pacific Conference Series, Vol. 336, Cosmic Abundances as Records of
  Stellar Evolution and Nucleosynthesis, ed. T.~G. {Barnes}, III \& F.~N.
  {Bash}, 25

\bibitem[{{Banse} {et~al.}(1992){Banse}, {Grosbol}, \&
  {Baade}}]{1992ASPC...25..120B}
{Banse}, K., {Grosbol}, P., \& {Baade}, D. 1992, in Astronomical Society of the
  Pacific Conference Series, Vol.~25, Astronomical Data Analysis Software and
  Systems I, ed. D.~M. {Worrall}, C.~{Biemesderfer}, \& J.~{Barnes}, 120

\bibitem[{{B{\'{\i}}r{\'o}} \& {Nuspl}(2011)}]{2011MNRAS.416.1601B}
{B{\'{\i}}r{\'o}}, I.~B. \& {Nuspl}, J. 2011, \mnras, 416, 1601

\bibitem[{{Breger} {et~al.}(1993){Breger}, {Stich}, {Garrido}, {Martin},
  {Jiang}, {Li}, {Hube}, {Ostermann}, {Paparo}, \&
  {Scheck}}]{1993A&A...271..482B}
{Breger}, M., {Stich}, J., {Garrido}, R., {et~al.} 1993, \aap, 271, 482

\bibitem[{{Budding}(1989)}]{1989SSRv...50..205B}
{Budding}, E. 1989, \ssr, 50, 205

\bibitem[{{Budding} \& {Butland}(2011, BB2011)}]{2011MNRAS.418.1764B}
{Budding}, E. \& {Butland}, R. 2011, BB2011, \mnras, 418, 1764

\bibitem[{{Deschamps} {et~al.}(2015){Deschamps}, {Braun}, {Jorissen}, {Siess},
  {Baes}, \& {Camps}}]{2015A&A...577A..55D}
{Deschamps}, R., {Braun}, K., {Jorissen}, A., {et~al.} 2015, \aap, 577, A55

\bibitem[{{Deschamps} {et~al.}(2013){Deschamps}, {Siess}, {Davis}, \&
  {Jorissen}}]{2013A&A...557A..40D}
{Deschamps}, R., {Siess}, L., {Davis}, P.~J., \& {Jorissen}, A. 2013, \aap,
  557, A40

\bibitem[{{Donati} {et~al.}(1997){Donati}, {Semel}, {Carter}, {Rees}, \&
  {Collier Cameron}}]{1997MNRAS.291..658D}
{Donati}, J.-F., {Semel}, M., {Carter}, B.~D., {Rees}, D.~E., \& {Collier
  Cameron}, A. 1997, \mnras, 291, 658

\bibitem[{{ESA}(1997)}]{1997ESASP1200.....E}
{ESA}, ed. 1997, ESA Special Publication, Vol. 1200, {The HIPPARCOS and TYCHO
  catalogues. Astrometric and photometric star catalogues derived from the ESA
  HIPPARCOS Space Astrometry Mission}

\bibitem[{{Gamarova} {et~al.}(2003){Gamarova}, {Mkrtichian}, {Rodriguez},
  {Costa}, \& {Lopez-Gonzalez}}]{2003ASPC..292..369G}
{Gamarova}, A.~Y., {Mkrtichian}, D.~E., {Rodriguez}, E., {Costa}, V., \&
  {Lopez-Gonzalez}, M.~J. 2003, in Astronomical Society of the Pacific
  Conference Series, Vol. 292, Interplay of Periodic, Cyclic and Stochastic
  Variability in Selected Areas of the H-R Diagram, ed. C.~{Sterken}, 369

\bibitem[{{Glazunova} {et~al.}(2009){Glazunova}, {Yushchenko}, \&
  {Mkrtichian}}]{Glazunova2009}
{Glazunova}, L.~V., {Yushchenko}, A.~V., \& {Mkrtichian}, D.~E. 2009,
  Kinematics Phys. Celest. Bodies (Ukraine), 6, 324

\bibitem[{{Guinan}(1977)}]{1977AJ.....82...51G}
{Guinan}, E.~F. 1977, \aj, 82, 51

\bibitem[{{Guo} {et~al.}(2016){Guo}, {Gies}, {Matson}, \& {Garc{\'{\i}}a
  Hern{\'a}ndez}}]{2016ApJ...826...69G}
{Guo}, Z., {Gies}, D.~R., {Matson}, R.~A., \& {Garc{\'{\i}}a Hern{\'a}ndez}, A.
  2016, \apj, 826, 69

\bibitem[{{Guo} {et~al.}(2017){Guo}, {Gies}, {Matson}, {Garc{\'{\i}}a
  Hern{\'a}ndez}, {Han}, \& {Chen}}]{2017ApJ...837..114G}
{Guo}, Z., {Gies}, D.~R., {Matson}, R.~A., {et~al.} 2017, \apj, 837, 114

\bibitem[{{Gustafsson} {et~al.}(2008){Gustafsson}, {Edvardsson}, {Eriksson},
  {J{\o}rgensen}, {Nordlund}, \& {Plez}}]{2008A&A...486..951G}
{Gustafsson}, B., {Edvardsson}, B., {Eriksson}, K., {et~al.} 2008, \aap, 486,
  951

\bibitem[{{Hadrava}(1995)}]{1995A&AS..114..393H}
{Hadrava}, P. 1995, \aaps, 114, 393

\bibitem[{{Hadrava}(2006)}]{Hadrava2006Ap&SS.304..337H}
{Hadrava}, P. 2006, \apss, 304, 337

\bibitem[{{Hensberge} {et~al.}(2008){Hensberge}, {Iliji{\'c}}, \&
  {Torres}}]{2008A&A...482.1031H}
{Hensberge}, H., {Iliji{\'c}}, S., \& {Torres}, K.~B.~V. 2008, \aap, 482, 1031

\bibitem[{{Kopal}(1956)}]{1956AnAp...19..298K}
{Kopal}, Z. 1956, Annales d'Astrophysique, 19, 298

\bibitem[{{Kupka} {et~al.}(2000){Kupka}, {Ryabchikova}, {Piskunov}, {Stempels},
  \& {Weiss}}]{2000BaltA...9..590K}
{Kupka}, F.~G., {Ryabchikova}, T.~A., {Piskunov}, N.~E., {Stempels}, H.~C., \&
  {Weiss}, W.~W. 2000, Baltic Astronomy, 9, 590

\bibitem[{{Kuschnig} {et~al.}(1997){Kuschnig}, {Weiss}, {Gruber}, {Bely}, \&
  {Jenkner}}]{1997A&A...328..544K}
{Kuschnig}, R., {Weiss}, W.~W., {Gruber}, R., {Bely}, P.~Y., \& {Jenkner}, H.
  1997, \aap, 328, 544

\bibitem[{{Lee} {et~al.}(2016){Lee}, {Kim}, {Hong}, {Koo}, {Lee}, \&
  {Youn}}]{2016AJ....151...25L}
{Lee}, J.~W., {Kim}, S.-L., {Hong}, K., {et~al.} 2016, \aj, 151, 25

\bibitem[{{Lehmann} \& {Mkrtichian}(2004)}]{2004A&A...413..293L}
{Lehmann}, H. \& {Mkrtichian}, D.~E. 2004, \aap, 413, 293

\bibitem[{{Lehmann} \& {Mkrtichian}(2008)}]{2008A&A...480..247L}
{Lehmann}, H. \& {Mkrtichian}, D.~E. 2008, \aap, 480, 247

\bibitem[{{Lehmann} {et~al.}(2013){Lehmann}, {Southworth}, {Tkachenko}, \&
  {Pavlovski}}]{2013A&A...557A..79L}
{Lehmann}, H., {Southworth}, J., {Tkachenko}, A., \& {Pavlovski}, K. 2013,
  \aap, 557, A79

\bibitem[{{Lenz} \& {Breger}(2005)}]{2005CoAst.146...53L}
{Lenz}, P. \& {Breger}, M. 2005, Communications in Asteroseismology, 146, 53

\bibitem[{{Liakos} {et~al.}(2012){Liakos}, {Niarchos}, {Soydugan}, \&
  {Zasche}}]{2012MNRAS.422.1250L}
{Liakos}, A., {Niarchos}, P., {Soydugan}, E., \& {Zasche}, P. 2012, \mnras,
  422, 1250

\bibitem[{{Maceroni} {et~al.}(2014){Maceroni}, {Lehmann}, {da Silva},
  {Montalb{\'a}n}, {Lee}, {Ak}, {Deshpande}, {Yakut}, {Debosscher}, {Guo},
  {Kim}, {Lee}, \& {Southworth}}]{2014A&A...563A..59M}
{Maceroni}, C., {Lehmann}, H., {da Silva}, R., {et~al.} 2014, \aap, 563, A59

\bibitem[{{McLaughlin}(1924)}]{1924ApJ....60...22M}
{McLaughlin}, D.~B. 1924, \apj, 60

\bibitem[{{Mennekens} {et~al.}(2008){Mennekens}, {De Greve}, {van Rensbergen},
  \& {Yungelson}}]{2008A&A...486..919M}
{Mennekens}, N., {De Greve}, J.-P., {van Rensbergen}, W., \& {Yungelson}, L.~R.
  2008, \aap, 486, 919

\bibitem[{{Mirtorabi} \& {Riazi}(2006)}]{2006Ap&SS.306..159M}
{Mirtorabi}, M.~T. \& {Riazi}, N. 2006, \apss, 306, 159

\bibitem[{{Mkrtichian} {et~al.}(2006){Mkrtichian}, {Kim}, {Kusakin},
  {Rovithis-Livaniou}, {Rovithis}, {Lampens}, {van Cauteren}, {Shobbrook},
  {Rodriguez}, {Gamarova}, {Olson}, \& {Kang}}]{2006Ap&SS.304..169M}
{Mkrtichian}, D., {Kim}, S.-L., {Kusakin}, A.~V., {et~al.} 2006, \apss, 304,
  169

\bibitem[{{Mkrtichian} \& {Gamarova}(2000)}]{2000IBVS.4836....1M}
{Mkrtichian}, D.~E. \& {Gamarova}, A.~Y. 2000, Information Bulletin on Variable
  Stars, 4836

\bibitem[{{Mkrtichian} {et~al.}(2002){Mkrtichian}, {Kusakin}, {Gamarova}, \&
  {Nazarenko}}]{2002ASPC..259...96M}
{Mkrtichian}, D.~E., {Kusakin}, A.~V., {Gamarova}, A.~Y., \& {Nazarenko}, V.
  2002, in Astronomical Society of the Pacific Conference Series, Vol. 259, IAU
  Colloq. 185: Radial and Nonradial Pulsationsn as Probes of Stellar Physics,
  ed. C.~{Aerts}, T.~R. {Bedding}, \& J.~{Christensen-Dalsgaard}, 96

\bibitem[{{Mkrtichian} {et~al.}(2018){Mkrtichian}, {Lehmann},
  {Rodr{\'{\i}}guez}, {Olson}, {Kim}, {Kusakin}, {Lee}, {Youn}, {Kwon},
  {L{\'o}pez-Gonz{\'a}lez}, {Janiashvili}, {Tiwari}, {Joshi}, {Lampens}, {Van
  Cauteren}, {Glazunova}, {Gamarova}, {Grankin}, {Rovithis-Livaniou},
  {Svoboda}, {Uhlar}, {Tsymbal}, {Kokumbaeva}, {Urushadze}, {Kuratov}, {Shin},
  {Kang}, \& {Soonthornthum}}]{2018MNRAS.475.4745M}
{Mkrtichian}, D.~E., {Lehmann}, H., {Rodr{\'{\i}}guez}, E., {et~al.} 2018,
  \mnras, 475, 4745

\bibitem[{{Mkrtichian} {et~al.}(2005){Mkrtichian}, {Rodr{\'{\i}}guez}, {Olson},
  {Kusakin}, {Kim}, {Lehmann}, {Gamarova}, \& {Kang}}]{2005ASPC..333..197M}
{Mkrtichian}, D.~E., {Rodr{\'{\i}}guez}, E., {Olson}, E.~C., {et~al.} 2005, in
  Astronomical Society of the Pacific Conference Series, Vol. 333, Tidal
  Evolution and Oscillations in Binary Stars, ed. A.~{Claret},
  A.~{Gim{\'e}nez}, \& J.-P. {Zahn}, 197

\bibitem[{{Mochnacki}(1984)}]{1984ApJS...55..551M}
{Mochnacki}, S.~W. 1984, \apjs, 55, 551

\bibitem[{{Paschke} \& {Brat}(2006)}]{2006OEJV...23...13P}
{Paschke}, A. \& {Brat}, L. 2006, Open European Journal on Variable Stars, 23,
  13

\bibitem[{{Pavlovski} \& {Hensberge}(2010)}]{2010ASPC..435..207P}
{Pavlovski}, K. \& {Hensberge}, H. 2010, in Astronomical Society of the Pacific
  Conference Series, Vol. 435, Binaries - Key to Comprehension of the Universe,
  ed. A.~{Pr{\v s}a} \& M.~{Zejda}, 207

\bibitem[{{Pr{\v s}a} {et~al.}(2016){Pr{\v s}a}, {Conroy}, {Horvat}, {Pablo},
  {Kochoska}, {Bloemen}, {Giammarco}, {Hambleton}, \&
  {Degroote}}]{2016ApJS..227...29P}
{Pr{\v s}a}, A., {Conroy}, K.~E., {Horvat}, M., {et~al.} 2016, \apjs, 227, 29

\bibitem[{{Pr{\v s}a} \& {Zwitter}(2005)}]{2005ApJ...628..426P}
{Pr{\v s}a}, A. \& {Zwitter}, T. 2005, \apj, 628, 426

\bibitem[{{Radhakrishnan} {et~al.}(1984){Radhakrishnan}, {Abhyankar}, \&
  {Sarma}}]{1984BASI...12..182R}
{Radhakrishnan}, K.~R., {Abhyankar}, K.~D., \& {Sarma}, M.~B.~K. 1984, Bulletin
  of the Astronomical Society of India, 12, 182

\bibitem[{{Radhakrishnan} {et~al.}(1985){Radhakrishnan}, {Sarma}, \&
  {Abhyankar}}]{1985BASI...13..261R}
{Radhakrishnan}, K.~R., {Sarma}, M.~B.~K., \& {Abhyankar}, K.~D. 1985, Bulletin
  of the Astronomical Society of India, 13, 261

\bibitem[{{Raskin} {et~al.}(2011){Raskin}, {van Winckel}, {Hensberge},
  {Jorissen}, {Lehmann}, {Waelkens}, {Avila}, {de Cuyper}, {Degroote},
  {Dubosson}, {Dumortier}, {Fr{\'e}mat}, {Laux}, {Michaud}, {Morren}, {Perez
  Padilla}, {Pessemier}, {Prins}, {Smolders}, {van Eck}, \&
  {Winkler}}]{2011A&A...526A..69R}
{Raskin}, G., {van Winckel}, H., {Hensberge}, H., {et~al.} 2011, \aap, 526, A69

\bibitem[{{Ribas} {et~al.}(2002){Ribas}, {Arenou}, \&
  {Guinan}}]{2002AJ....123.2033R}
{Ribas}, I., {Arenou}, F., \& {Guinan}, E.~F. 2002, \aj, 123, 2033

\bibitem[{{Rodr{\'{\i}}guez} {et~al.}(2004{\natexlab{a}}){Rodr{\'{\i}}guez},
  {Garc{\'{\i}}a}, {Gamarova}, {Costa}, {Daszy{\'n}ska-Daszkiewicz},
  {L{\'o}pez-Gonz{\'a}lez}, {Mkrtichian}, \& {Rolland}}]{2004MNRAS.353..310R}
{Rodr{\'{\i}}guez}, E., {Garc{\'{\i}}a}, J.~M., {Gamarova}, A.~Y., {et~al.}
  2004{\natexlab{a}}, \mnras, 353, 310

\bibitem[{{Rodr{\'{\i}}guez} {et~al.}(2004{\natexlab{b}}){Rodr{\'{\i}}guez},
  {Garc{\'{\i}}a}, {Mkrtichian}, {Costa}, {Kim}, {L{\'o}pez-Gonz{\'a}lez},
  {Hintz}, {Kusakin}, {Gamarova}, {Lee}, {Youn}, {Janiashvili}, {Garrido},
  {Moya}, \& {Kang}}]{2004MNRAS.347.1317R}
{Rodr{\'{\i}}guez}, E., {Garc{\'{\i}}a}, J.~M., {Mkrtichian}, D.~E., {et~al.}
  2004{\natexlab{b}}, \mnras, 347, 1317

\bibitem[{{Rossiter}(1924)}]{1924ApJ....60...15R}
{Rossiter}, R.~A. 1924, \apj, 60

\bibitem[{{Sarma} {et~al.}(1996){Sarma}, {Rao}, \&
  {Abhyankar}}]{1996ApJ...458..371S}
{Sarma}, M.~B.~K., {Rao}, P.~V., \& {Abhyankar}, K.~D. 1996, \apj, 458, 371

\bibitem[{{Sarna} \& {Ziolkowski}(1988)}]{1988AcA....38...89S}
{Sarna}, M.~J. \& {Ziolkowski}, J. 1988, \actaa, 38, 89

\bibitem[{{Sato}(1971)}]{1971PASJ...23..335S}
{Sato}, K. 1971, \pasj, 23, 335

\bibitem[{{Shulyak} {et~al.}(2004){Shulyak}, {Tsymbal}, {Ryabchikova},
  {St{\"u}tz}, \& {Weiss}}]{2004A&A...428..993S}
{Shulyak}, D., {Tsymbal}, V., {Ryabchikova}, T., {St{\"u}tz}, C., \& {Weiss},
  W.~W. 2004, \aap, 428, 993

\bibitem[{{Soydugan} {et~al.}(2006){Soydugan}, {Soydugan}, {Ibano{\u g}lu},
  {Frasca}, {Demircan}, \& {Akan}}]{2006AN....327..905S}
{Soydugan}, E., {Soydugan}, F., {Ibano{\u g}lu}, C., {et~al.} 2006,
  Astronomische Nachrichten, 327, 905

\bibitem[{{Sterne}(1941)}]{1941PNAS...27..175S}
{Sterne}, T.~E. 1941, Proceedings of the National Academy of Science, 27, 175

\bibitem[{Tkachenko(2010)}]{2010Tkachenko}
Tkachenko, A. 2010, PhD Thesis, Friedrich-Schiller-Universit\"at Jena,
  http://www.thulb.uni-jena.de/Online\_Katalog.html

\bibitem[{{Tkachenko}(2015)}]{2015A&A...581A.129T}
{Tkachenko}, A. 2015, \aap, 581, A129

\bibitem[{{Tkachenko} {et~al.}(2009){Tkachenko}, {Lehmann}, \&
  {Mkrtichian}}]{2009A&A...504..991T}
{Tkachenko}, A., {Lehmann}, H., \& {Mkrtichian}, D.~E. 2009, \aap, 504, 991

\bibitem[{{Tkachenko} {et~al.}(2013){Tkachenko}, {Van Reeth}, {Tsymbal},
  {Aerts}, {Kochukhov}, \& {Debosscher}}]{2013A&A...560A..37T}
{Tkachenko}, A., {Van Reeth}, T., {Tsymbal}, V., {et~al.} 2013, \aap, 560, A37

\bibitem[{{Tomkin}(1985)}]{1985ApJ...297..250T}
{Tomkin}, J. 1985, \apj, 297, 250

\bibitem[{{Tomkin} \& {Lambert}(1989)}]{1989MNRAS.241..777T}
{Tomkin}, J. \& {Lambert}, D.~L. 1989, \mnras, 241, 777

\bibitem[{{Tsymbal}(1996)}]{1996ASPC..108..198T}
{Tsymbal}, V. 1996, in Astronomical Society of the Pacific Conference Series,
  Vol. 108, M.A.S.S., Model Atmospheres and Spectrum Synthesis, ed. S.~J.
  {Adelman}, F.~{Kupka}, \& W.~W. {Weiss}, 198

\bibitem[{{{\v S}koda} \& {Hadrava}(2010)}]{2010ASPC..435...71S}
{{\v S}koda}, P. \& {Hadrava}, P. 2010, in Astronomical Society of the Pacific
  Conference Series, Vol. 435, Binaries - Key to Comprehension of the Universe,
  ed. A.~{Pr{\v s}a} \& M.~{Zejda}, 71

\bibitem[{{{\v S}koda} {et~al.}(2012){{\v S}koda}, {Hadrava}, \&
  {Fuchs}}]{2012IAUS..282..403S}
{{\v S}koda}, P., {Hadrava}, P., \& {Fuchs}, J. 2012, in IAU Symposium, Vol.
  282, From Interacting Binaries to Exoplanets: Essential Modeling Tools, ed.
  M.~T. {Richards} \& I.~{Hubeny}, 403--404

\bibitem[{{van Rensbergen} {et~al.}(2008){van Rensbergen}, {De Greve}, {De
  Loore}, \& {Mennekens}}]{2008A&A...487.1129V}
{van Rensbergen}, W., {De Greve}, J.~P., {De Loore}, C., \& {Mennekens}, N.
  2008, \aap, 487, 1129

\bibitem[{{van Rensbergen} {et~al.}(2010){van Rensbergen}, {De Greve},
  {Mennekens}, {Jansen}, \& {De Loore}}]{2010A&A...510A..13V}
{van Rensbergen}, W., {De Greve}, J.~P., {Mennekens}, N., {Jansen}, K., \& {De
  Loore}, C. 2010, \aap, 510, A13

\bibitem[{{Varricatt} \& {Ashok}(1999)}]{1999AJ....117.2980V}
{Varricatt}, W.~P. \& {Ashok}, N.~M. 1999, \aj, 117, 2980

\bibitem[{{Webbink}(1977)}]{1977ApJ...211..486W}
{Webbink}, R.~F. 1977, \apj, 211, 486

\bibitem[{{Wilson}(1979)}]{1979ApJ...234.1054W}
{Wilson}, R.~E. 1979, \apj, 234, 1054

\bibitem[{{Wood}(1946)}]{1946CoPri...22..1W}
{Wood}, F.~B. 1946, Contributions from the Princeton University Observatory,
  22, 1

\bibitem[{{Zima}(2008)}]{2008CoAst.157..387Z}
{Zima}, W. 2008, Communications in Asteroseismology, 157, 387

\bibitem[{{Zucker} \& {Mazeh}(1994)}]{1994ApJ...420..806Z}
{Zucker}, S. \& {Mazeh}, T. 1994, \apj, 420, 806

\end{thebibliography}

\end{document}